\documentclass[reprint,aps,prb,groupedaddress,showpacs,fleqn]{revtex4-1}
\usepackage[latin9]{inputenc}
\usepackage{verbatim}
\usepackage{rotfloat}
\usepackage{amsmath}
\usepackage{amssymb}
\usepackage{graphicx}
\usepackage{esint}
\usepackage{ulem}

\makeatletter

\providecommand{\tabularnewline}{\\}

 \@ifundefined{textcolor}{}
 {%
   \definecolor{BLACK}{gray}{0}
   \definecolor{WHITE}{gray}{1}
   \definecolor{RED}{rgb}{1,0,0}
   \definecolor{GREEN}{rgb}{0,1,0}
   \definecolor{BLUE}{rgb}{0,0,1}
   \definecolor{CYAN}{cmyk}{1,0,0,0}
   \definecolor{MAGENTA}{cmyk}{0,1,0,0}
   \definecolor{YELLOW}{cmyk}{0,0,1,0}
 }

\usepackage[raggedright]{titlesec}\usepackage{lipsum}\usepackage{colortbl}\usepackage{gensymb}
\usepackage{color}\usepackage{soul}\usepackage{rotating}

\makeatother

\begin{document}

\title{Identifying and tracing potential energy surfaces of electronic excitations 
with specific character via their transition origins: application to oxirane}

\author{Jian-Hao Li}
\email{jhl52@cam.ac.uk}


\author{T. J. Zuehlsdorff}


\author{M. C. Payne}


\author{N. D. M. Hine}

\affiliation{TCM Group, Cavendish Laboratory, J. J. Thomson Avenue, Cambridge CB3 0HE, United
Kingdom}

\date{\today}
\begin{abstract}
We show that the transition origins of electronic excitations identified
by quantified natural transition orbital (QNTO) analysis can be employed
to connect potential energy surfaces (PESs) according to their character across a wide
range of molecular geometries. This is achieved by locating the switching of 
transition origins 
of adiabatic potential surfaces as the geometry changes. The transition vectors
for analysing transition origins are provided by linear response time-dependent
density functional theory (TDDFT) calculations under the Tamm-Dancoff
approximation. We study the photochemical CO ring opening of
oxirane as an example and show that the results corroborate the traditional
Gomer-Noyes mechanism derived experimentally. The knowledge of specific states
for the reaction also agrees well with that given by previous theoretical
work using TDDFT surface-hopping dynamics that was validated by high-quality
quantum Monte Carlo calculations.
We also show that QNTO can be useful for considerably larger and more complex
systems: by projecting the excitations to those of a reference
oxirane molecule, the approach is able to identify and analyse specific
excitations of a trans-2,3-diphenyloxirane molecule.
\end{abstract}
\maketitle

\section{Introduction}

Kohn-Sham (KS) density functional theory (DFT)\cite{HK64,KS65,PY94,GV05}
can accurately reproduce many properties of molecular and solid state
systems\cite{KBP96,B97,CH08} as compared to experimental methods
and higher-level theory, e.g.~molecular structures, vibrational frequencies,
atomisation energies and ionisation potentials. Meanwhile, its time-dependent
extension, TDDFT\cite{RG84,MUN06,C95,DH05}, particularly in the
linear response (LR) formalism\cite{C95}, can deliver properties
of excited states and optical absorption spectra of molecular systems.
DFT and TDDFT calculations are increasingly able to treat large, complex
biological systems, such as proteins, enzymes, and nucleic acids,
due to the increasing availability of large computational facilities in
recent years\cite{LC14,CC13,FP13,TB13}. However, there remain serious obstacles to
such uses: the intrinsic cubic scaling of traditional KS-DFT with
respect to the number of electrons; the explosion of the size of the
available configurational phase space associated with large, flexible
molecules; and the challenges of accurate electrostatics of solvated
systems\cite{LCHHP}. When excited state properties are considered,
further challenges come into play such as the proliferation of spurious
charge transfer excitations, and the difficulty of
identifying connected
excited states on a PES.

Recent advances in linear-scaling (LS)-DFT\cite{WH94,WC06,SSF96,HG95,FG06,VK05,SA02,SHMP05}
have made the study of very large molecular systems much more feasible.
LS-DFT relies on the exponential decay of the density matrix, $\rho(\mathbf{r,r'})$,
with respect to $|\mathbf{r-r'}|$ in a system with a finite band-gap\cite{BH97,IA99,HV01}.
In such systems, the density matrix, expressed using local orbitals,
can be truncated by value or by range to render it local. In this
work we use the ONETEP code\cite{SHMP05}, where a minimal number
of local optimisable functions, called nonorthogonal generalised Wannier
functions (NGWFs), are optimised in situ to best represent the density
matrix of the system. Recently, Linear-Response TDDFT Tamm-Dancoff
approximation (TDA) calculations have also become available in ONETEP,
using a similar density matrix formalism\cite{ZH13}. Such methods
can provide valuable knowledge about the electronic excitations of very
large molecular systems such as proteins and DNA segments. However,
the low symmetry and complex geometry of biological systems exacerbates
the aforementioned problem of tracing the PES
of specific excited states. This work therefore focuses on quantified
natural transition orbital (QNTO) analysis, recently proposed by Li
\textit{et al.}\cite{LCGH11}, which offers progress towards this goal.

QNTO analysis rewrites the transition vectors of an electronic excitation
identified by a theoretical calculation, e.g. TDDFT, in terms of the natural
transition orbitals (NTOs)\cite{M03,DH05,PWD14}.
The projection of hole (NTO$n$-H) and electron (NTO$n$-E) orbitals of the $n$th
NTO pair onto a standard orbital basis set such as natural bond orbitals
(NBOs)\cite{RCW88,GLW12,LC12} then helps quantify the transition
origins of the NTOs of the excitation. Alongside the absorption energy
and oscillator strength, knowledge of these transition origins helps provide
a physical interpretation of the role of a given excitation dominated by single 
excitation character.  Similar analysis
may also be developed for single de-excitations, which is not as
important in generic biological systems where de-excitations are usually 
negligible\cite{DH05}, signified by the good approximation of TDDFT/TDA to TDDFT.
In addition, even in large and
complex systems, electronic excitations may involve numerous electronic
transitions between valence and conduction orbitals; using NTOs as
the orbital basis, the number of dominant electronic transitions is
often reduced to one or two, helping clarify what transition components
contribute to the electronic excitations. QNTO analysis can help to
assess the variation of transition origins determined by the different theoretical
methods available in an electronic excitation calculation. For example, this
approach has been used to assess the performance of different exchange-correlation
functionals in predicting transition origins by comparing to a high-level 
theoretical calculation\cite{LCGH11}.  
The variations of transition origins caused by the presence of different
molecular environments and/or different molecular geometries can also
be investigated. For example, QNTO has been used to study the role
of the DNA backbone in mediating the transition origins of electronic
excitations of B-DNA\cite{LCGH12}.

In the present paper, we introduce theoretical background to Linear-Response
calculations in ONETEP (Sec. II), and show that QNTO analysis can
be implemented with a non-orthogonal basis set such as the NGWFs used
in ONETEP (Sec. III). In Sec. IV we then demonstrate two applications
of LR-TDDFT/TDA/QNTO: (1) studying the photo-driven ring opening reaction
of oxirane, and (2) identifying and comparing similar excitations
between trans-2,3-diphenyloxirane and oxirane. We demonstrate that
QNTO analysis can be readily used to locate state crossings that are
otherwise unclear from plotting potential energy curves. The results
are consistent with previous experimental and theoretical studies,
showing that the present method is a feasible and reliable approach
for investigating excited state related reactions.

\section{LR-TDDFT in ONETEP}

Linear-scaling DFT as implemented in ONETEP\cite{SMHDP02,SHMP05}
relies on a reformulation of the single-electron density matrix in
terms of local orbitals, as follows 
\begin{equation}
\begin{split}\rho(\mathbf{r,r'})= & \sum\limits _{p}f_{p}\psi_{p}(\mathbf{r})\psi_{p}^{\ast}(\mathbf{r'})\;\;\;\\
= & \phi_{\alpha}(\mathbf{r})K^{\alpha\beta}\phi_{\beta}^{\ast}(\mathbf{r'})
\end{split}
\end{equation}
where $\psi_{p}(\mathbf{r})$ are single-particle eigenstates with
occupation $f_{p}$, while $\phi_{\alpha}(\mathbf{r})$ are spatially-localised
nonorthogonal orbitals. $K^{\alpha\beta}$ is a generalised matrix
representation of the occupancies, known as the density kernel. Note
that implicit summation over repeated Greek indices is used throughout
this work. The density kernel is defined by $K^{\alpha\beta}=\langle\phi^{\alpha}|\hat{\rho}_{\mathrm{KS}}|\phi^{\beta}\rangle$,
with $\hat{\rho}_{\mathrm{KS}}$ being the KS single-particle density
operator. The nonorthogonal orbitals $\{\phi_{\alpha}\}$ and their
duals $\{\phi^{\alpha}\}$ form a biorthonormal basis set such that
$\langle\phi_{\alpha}|\phi^{\beta}\rangle=\delta_{\alpha}^{\beta}$.

ONETEP employs a minimal number of NGWFs, each expressed in terms
of periodic bandwidth limited delta functions (psincs) centred on
the grid points of the simulation cell, with a spacing determined
by a kinetic energy cutoff akin to that in a plane-wave code.\cite{SHMP5757}
NGWFs are localised to atom-centred spherical regions defined by a
cutoff $R_{\phi}$. A total energy calculation proceeds by minimising
the total energy simultaneously with respect to the matrix elements
of the density kernel and the expansion coefficients of the NGWFs
in terms of the psinc functions, via nested loops of conjugate gradients
optimisations\cite{MHSP03}. NGWFs are initialised to pseudoatomic
orbitals which solve the pseudopotential for a free atom.\cite{RHS12} We confine
our attention to closed-shell systems (even number of electrons) with
non-fractional occupancy, and the total electronic energy of the real
system is written as 
\begin{equation}
E_{0}=2\text{Tr}\left[\mathbf{K}_{\{v\}}\mathbf{H}^{\{v\}}\right]+E_{\textrm{DC}}
\end{equation}
where $\{v\}$ denotes `valence', $H_{\alpha\beta}^{\{v\}}=\langle\phi_{\alpha}|\hat{H}_{\mathrm{KS}}|\phi_{\beta}\rangle$
and $K_{\{v\}}^{\alpha\beta}=\langle\phi^{\alpha}|\hat{\rho}^{\{v\}}|\phi^{\beta}\rangle$
with $\hat{\rho}^{\{v\}}=\sum\limits _{v}^{\mathrm{occ}}|\psi_{v}\rangle\langle\psi_{v}|$.
The factor $2$ accounts for the spin degeneracy and $E_{\textrm{DC}}$
is the standard ``double-counting'' correction. An FFT box technique\cite{SMHDP02}
is employed to allow one to construct matrix elements and density
components in $O(1)$ computational effort independent of system size.

As described in detail in Ref. \onlinecite{HSMP08}, the minimisation
is carried under the constraint of conserved total electron number,
\begin{equation}
N=2\text{Tr}\left[\mathbf{K}_{\{v\}}\mathbf{S}^{\{v\}}\right]
\end{equation}
where $S_{\alpha\beta}^{\{v\}}=\langle\phi_{\alpha}|\phi_{\beta}\rangle$,
and also the constraint of idempotent density matrix, 
\begin{equation}
\textbf{K}_{\{v\}}=\textbf{K}_{\{v\}}\textbf{S}^{\{v\}}\textbf{K}_{\{v\}}
\end{equation}
which is equivalent to requiring the orthonormal orbitals have occupancy
of $0$ or $1$. An important advantage of optimising the NGWFs in
situ is that basis set superposition error (BSSE) is eliminated.\cite{HSMP06}
Sophisticated algorithms have been implemented so that a high performance
can be achieved running on parallel computers\cite{HHMSP09}.

Because the NGWFs are optimised in situ to describe the occupied part
of the density matrix, they do not in general provide a good description
of unoccupied states. Therefore, we instead generate and optimise
in situ a second set of NGWFs, denoted as $\{\chi_{\alpha}\}$, obtained
by minimising the bandstructure energy associated with a chosen range
of low-energy conduction band states, having first projected out the
valence states. The bandstructure energy to be minimised is written
as 
\begin{equation}
E=2\text{Tr}\left[\mathbf{K}_{\{c\}}\mathbf{H}^{\{c\}}\right]
\end{equation}
where $\{c\}$ denotes `conduction', $H_{\alpha\beta}^{\{c\}}=\langle\chi_{\alpha}|\hat{H}_{\mathrm{proj}}|\chi_{\beta}\rangle$
and $K_{\{c\}}^{\alpha\beta}=\langle\chi^{\alpha}|\hat{\rho}^{\{c\}}|\chi^{\beta}\rangle$.
A full description of this procedure can be found in Ref. \onlinecite{RHH11}.

Recently Zuehlsdorff \textit{et al.} proposed a formulation for linear-scaling
LR-TDDFT/TDA (hereafter simply denoted as TDDFT) calculations, which
uses the conduction and valence NGWFs as a basis for describing the
response densities of electronic excitations\cite{ZH13}. This approach
adapts the Casida formulation for finding transition vectors, $\mathbf{X}_{I}$,
and excitation energies $\omega_{I}$ by using the TDDFT equation,
$\mathbf{AX}_{I}=\omega_{I}\mathbf{X}_{I}$, to a nonorthogonal local
orbital basis. The quantity sought is the sum of the energies of the
$N_{\omega}$ lowest excitations
\begin{equation}
\Omega=\sum\limits _{I=1}^{N_{\omega}}\omega_{I}=\sum\limits _{I=1}^{N_{\omega}}\sum\limits _{cv,c'v'}x_{cv}^{I\ast}A_{cv,c'v'}x_{c'v'}^{I}\label{eq:Omega}
\end{equation}
A conjugate gradients minimisation of $\Omega$
is performed subject to the constraint of keeping the eigenvectors
orthonormal, expressing each transition $I$ using a transition density
kernel $\mathbf{K}_{\{1\}I}$ 
\[
\begin{split}\omega_{I} & =\text{Tr}\Bigg[\textbf{K}_{\{1\}I}^{\dagger}\textbf{S}^{\{c\}}\Bigg(\textbf{K}_{\{c\}}\textbf{H}^{\{c\}}\textbf{K}_{\{1\}I}-\\
 & \textbf{K}_{\{1\}I}\textbf{H}^{\{v\}}\textbf{K}_{\{v\}}+\textbf{K}_{\{c\}}\textbf{V}_{\mathrm{SCF}}^{\{1\}}\textbf{K}_{\{v\}}\Bigg)\textbf{S}^{\{v\}}\Bigg]
\end{split}
\;.
\]
The transition density kernel, $K_{\{1\}I}^{\alpha\beta}=\langle\chi^{\alpha}|\hat{\rho}^{\{1\}}|\phi^{\beta}\rangle$
with $\hat{\rho}^{\{1\}}=\sum_{cv}|\psi_{c}\rangle x_{cv}^{I}\langle\psi_{v}|$,
defines the transition density according to 
\begin{equation}
\rho^{\{1\}}(\mathbf{r})=\chi_{\alpha}(\mathbf{r})K_{\{1\}I}^{\alpha\beta}\phi_{\beta}^{\ast}(\mathbf{r})\;,
\end{equation}
and the Hartree-exchange-correlation self-consistent field matrix
is defined by:
\[
\left(V_{\mathrm{SCF}}^{\{1\}}\right){}_{\alpha\beta}=\int\,\mathrm{d}^{3}r \; \chi_{\alpha}^{\ast}(\mathbf{r})V_{\mathrm{SCF}}^{\{1\}}(\mathbf{r})\phi_{\beta}(\mathbf{r})
\]
with $V_{\mathrm{SCF}}^{\{1\}}(\mathbf{r})$ being the Hartree and
exchange-correlation response to the transition density as defined in Eq. 9 of Ref.
\onlinecite{ZH13}.

The minimisation of Eq. \ref{eq:Omega} proceeds as described in Ref.
\onlinecite{ZH13}.
To avoid unwanted valence-valence transitions, we project out these
unwanted components by constructing the kernel as
\begin{equation}
\textbf{K}_{\{1\}I}^{\prime}=\textbf{K}_{\{c\}}\textbf{S}^{\{c\}}\textbf{K}_{\{1\}I}\textbf{S}^{\{v\}}\textbf{K}_{\{v\}}\label{eq:proj}
\end{equation}

Additionally, we define a combined NGWF set by amalgamating valence
and conduction NGWF sets to create a joint set, $\{\zeta_{\alpha}\}=\{\phi_{\alpha}\}\cup\{\chi_{\alpha}\}$.
Using this representation, and the corresponding overlap matrix $\mathbf{S}^{\{j\}}$,
in place of $\{\chi_{\alpha}\}$ and $\mathbf{S}^{\{c\}}$, means
that we no longer need the explicit representation of the conduction
space density operator. Instead, the conduction space projector can
be replaced by subtracting the valence density operator from the identity
operator within the subspace of the full joint NGWF representation.
This defines $\textbf{K}_{\{j\}}$ through $K_{\{j\}}^{\alpha\beta}=\langle\zeta^{\alpha}|\hat{\rho}^{\{c\}}|\zeta^{\beta}\rangle=\langle\zeta^{\alpha}|\hat{I}-\hat{\rho}^{\{v\}}|\zeta^{\beta}\rangle$,
giving 
\begin{equation}
\textbf{K}_{\{j\}}=\textbf{S}_{\{j\}}^{-1}-\textbf{S}_{\{j\}}^{-1}\textbf{S}^{\{jv\}}\textbf{K}_{\{v\}}\textbf{S}^{\{jv\}\dagger}\textbf{S}_{\{j\}}^{-1}
\end{equation}
where $S_{\alpha\beta}^{\{jv\}}=\langle\zeta_{\alpha}|\phi_{\beta}\rangle$
and $\mathbf{S}_{\{j\}}^{-1}$ is the inverse of $\mathbf{S}^{\{j\}}$.

We also note that in the TDDFT the KS-system is linked to the real-system
by 
\begin{equation}
\sum_{cv}\psi_{v}^{\ast}(\mathbf{r})\psi_{c}(\mathbf{r})\langle\Psi_{0}|\hat{a}_{v}^{\dagger}\hat{a}_{c}|\Psi_{I}\rangle=\sum_{cv}\psi_{v}^{\ast}(\mathbf{r})\psi_{c}(\mathbf{r})x_{cv}^{I}
\end{equation}
where $|\Psi_{0}\rangle$ and $|\Psi_{I}\rangle$ are the `exact'
many-body wavefunctions that are obtained from diagonalising the exact
Hamiltonian in the configuration space.
The linear dependence among the products $\{\psi_{v}^{\ast}(\mathbf{r})\psi_{c}(\mathbf{r})\}$
becomes likely as the orbital basis approaches completion \cite{H86}
and the simple assignment $\langle\Psi_{0}|\hat{a}_{v}^{\dagger}\hat{a}_{c}|\Psi_{I}\rangle=x_{cv}^{I}$
may not be justified.
Nevertheless, this assignment would become reasonable\cite{C95} if
the number of orbitals involved in $\langle\Psi_{0}|\hat{a}_{v}^{\dagger}\hat{a}_{c}|\Psi_{I}\rangle$
is not too large, e.g. for low-lying excitations, and therefore the
study of transition origins becomes possible via directly analysing
$x_{cv}^{I}$ or $K_{\{1\}I}^{\alpha\beta}$ from the TDDFT calculation.

\section{Quantified natural transition orbital (QNTO) analysis}

In this section, we first introduce the basic formulation of QNTO
analysis in terms of an orthonormal orbital basis such as KS-orbitals.
We then detail how it is implemented in ONETEP within the NGWF
representation. The standard notation `$\dagger$' is used throughout to
denote conjugate transpose.

\subsection{Natural transition orbitals}

The NTO basis 
\begin{equation}
\begin{split}|\psi_{n}^{\{Nc\}}\rangle=\sum_{c}U_{cn}|\psi_{c}\rangle\\
|\psi_{n}^{\{Nv\}}\rangle=\sum_{v}V_{vn}|\psi_{v}\rangle
\end{split}
\end{equation}
is an orbital basis that makes $\mathbf{D}=\mathbf{U}^{\dagger}\mathbf{X}_{I}\mathbf{V}$
diagonal, with non-negative real numbers $\sqrt{\lambda_{n}}$ on
the diagonal. In this representation the transition density operator
is simplified to require summation over just one index: 
\begin{equation}
\hat{\rho}_{I}^{\{1\}}=\sum_{n=1}^{N_{N}}|\psi_{n}^{\{Nc\}}\rangle\sqrt{\lambda_{n}}\langle\psi_{n}^{\{Nv\}}|\;.
\end{equation}
$N_{N}$ denotes the number of non-zero values $\sqrt{\lambda_{n}}$,
which are ranked by magnitude. $\sqrt{\lambda_{n}}$ are called the
singular values from the singular value decomposition (SVD) of $\mathbf{X}_{I}=\mathbf{U}\mathbf{D}\mathbf{V}^{\dagger}$.
We also see that the $\mathbf{U}$ and $\mathbf{V}$ respectively
diagonalise $\mathbf{X}_{I}\mathbf{X}_{I}^{\dagger}$ and $\mathbf{X}_{I}^{\dagger}\mathbf{X}_{I}$,
\begin{equation}
\begin{split}\mathbf{U}^{\dagger}\mathbf{X}_{I}\mathbf{X}_{I}^{\dagger}\mathbf{U}=\mathbf{D}\mathbf{D}^{\dagger}\\
\mathbf{V}^{\dagger}\mathbf{X}_{I}^{\dagger}\mathbf{X}_{I}\mathbf{V}=\mathbf{D}^{\dagger}\mathbf{D}
\end{split}
\label{eq:NTOHE}
\end{equation}
with the same eigenvalue set $\{\lambda_{n}\}$, where $\mathbf{X}_{I}\mathbf{X}_{I}^{\dagger}$ and $\mathbf{X}_{I}^{\dagger}\mathbf{X}_{I}$
are density matrices for the electron and hole orbitals, respectively.

In TDDFT calculations in ONETEP, the local orbitals (NGWFs) are not
an orthonormal basis. If the NTOs are expressed in terms of NGWFs
as 
\begin{equation}
\begin{split}|\psi_{n}^{\{Nc\}}\rangle=|\zeta_{\alpha}\rangle\overline{U}_{\; n}^{\alpha}\\
|\psi_{n}^{\{Nv\}}\rangle=|\phi_{\alpha}\rangle\overline{V}_{\; n}^{\alpha}
\end{split}
\label{eq:NTOinNGWFs}
\end{equation}
$\mathbf{X}_{I}$ is then rewritten as $\mathbf{X}_{I}=\mathbf{U}\mathbf{\overline{U}}^{\dagger}\mathbf{S}^{\{c\}}\mathbf{K}_{\{1\}}\mathbf{S}^{\{v\}}\mathbf{\overline{V}}\mathbf{V}^{\dagger}$
and the target for decomposition becomes 
\begin{equation}
\mathbf{K}_{\{1\}}=\mathbf{\overline{U}}\mathbf{D}\mathbf{\overline{V}}^{\dagger}\label{eq:SVDinNGWFs}
\end{equation}
If we hope to directly employ an SVD algorithm for orthonormal basis,
we would have to first construct 
$\mathbf{M}\mathbf{K}_{\{1\}}\mathbf{N}^{\dagger}=(\mathbf{M}\mathbf{\overline{U}})\mathbf{D}(\mathbf{N}\mathbf{\overline{V}})^{\dagger}$,
where $\mathbf{M}^{\dagger}\mathbf{M}=\mathbf{S}^{\{c\}}$ and $\mathbf{N}^{\dagger}\mathbf{N}=\mathbf{S}^{\{v\}}$.
Once the singular vector matrices $\mathbf{M}\mathbf{\overline{U}}$
and $\mathbf{N}\mathbf{\overline{V}}$ are obtained, we would then
need to calculate $\mathbf{M}^{-1}$ and $\mathbf{N}^{-1}$ in order
to solve for $\mathbf{\overline{U}}$ and $\mathbf{\overline{V}}$.

However, since the square root and inverse of a sparse matrix are
in general much less sparse, their calculation would be very costly
when dealing with large matrices. As an alternative approach, we can
instead choose to diagonalise the $\mathbf{X}_{I}\mathbf{X}_{I}^{\dagger}$
and $\mathbf{X}_{I}^{\dagger}\mathbf{X}_{I}$ as shown in Eq. \ref{eq:NTOHE}.
Since $\mathbf{\overline{V}}^{\dagger}\mathbf{S}^{\{v\}}=\mathbf{\overline{V}}^{-1}$
and $\mathbf{\overline{U}}^{\dagger}\mathbf{S}^{\{c\}}=\mathbf{\overline{U}}^{-1}$,
the eigenvalue equations can be derived as 
\begin{equation}
\begin{split}(\mathbf{S}^{\{c\}}\mathbf{K}_{\{1\}}\mathbf{S}^{\{v\}}\mathbf{K}_{\{1\}}^{\dagger}\mathbf{S}^{\{c\}})\mathbf{\overline{U}}=(\mathbf{D}\mathbf{D}^{\dagger})\mathbf{S}^{\{c\}}\mathbf{\overline{U}}\\
(\mathbf{S}^{\{v\}}\mathbf{K}_{\{1\}}^{\dagger}\mathbf{S}^{\{c\}}\mathbf{K}_{\{1\}}\mathbf{S}^{\{v\}})\mathbf{\overline{V}}=(\mathbf{D}^{\dagger}\mathbf{D})\mathbf{S}^{\{v\}}\mathbf{\overline{V}}
\end{split}
\label{eq:NTOHEinNGWFs}
\end{equation}

Note that in the SVD of $\mathbf{X}_{I}=\mathbf{U}\mathbf{D}\mathbf{V}^{\dagger}$,
the relative phase between each pair of left (NTO$n$-E) and right
(NTO$n$-H) singular vectors are fixed: if the left vector is multiplied
by $(-1)$, the right vector must be multiplied by $(-1)$ to remain a
solution. Hence, the relative phase between each pair of eigenvectors
needs to be determined after solving Eq. \ref{eq:NTOHEinNGWFs}.
The following relation from Eq. \ref{eq:SVDinNGWFs} can be used: 
\begin{equation}
K_{\{1\}}^{\alpha\beta}=\sum_{n=1}^{N_{N}}\sqrt{\lambda_{n}}\overline{U}_{\; n}^{\alpha}\overline{V}_{\; n}^{\beta}
\label{eq:K1defviaUV}
\end{equation}
We define two matrices $\mathbf{d_{\textrm{n}+}}$ and $\mathbf{d_{\textrm{n}-}}$ as 
\begin{equation}
d_{\textrm{n}\pm}^{\alpha\beta}=K_{\{1\}}^{\alpha\beta}-\left(\sum_{m=1}^{n-1}\sqrt{\lambda_{m}}\overline{U}_{\; m}^{\alpha}\overline{V}_{\; m}^{\beta}\pm\sqrt{\lambda_{n}}\overline{U}_{\; n}^{\alpha}\overline{V}_{\; n}^{\beta}\right)
\label{eq:dpmdefviaKUV}
\end{equation}
From Eqs. \ref{eq:K1defviaUV} and \ref{eq:dpmdefviaKUV}, we see there is the relation: 
\begin{equation}
\textrm{Tr}\left[\mathbf{d}_{\textrm{n}+}\mathbf{S}^{\{v\}}\mathbf{d}_{\textrm{n}+}^{\dagger}\mathbf{S}^{\{c\}}\right]<\textrm{Tr}\left[\mathbf{d}_{\textrm{n}-}\mathbf{S}^{\{v\}}\mathbf{d}_{\textrm{n}-}^{\dagger}\mathbf{S}^{\{c\}}\right]\label{eq:DevNorm}
\end{equation}
We can thus start from NTO$1$ and proceed iteratively through higher NTOs,
checking which of the associated two eigenvectors obtained in Eq. \ref{eq:NTOHEinNGWFs},
denoted as $\overline{U}_{\; n}^{\prime\alpha}$ and $\overline{V}_{\; n}^{\prime\beta}$,
produces the smaller value in Eq. \ref{eq:DevNorm}. If Eq. \ref{eq:DevNorm} is not
obeyed, we can multiply the whole vector $\overline{V}_{\;1}^{\prime\beta}$ (or
$\overline{U}_{\;1}^{\prime\alpha}$) by $(-1)$ to obtain a correct
pair of $\overline{V}_{\;1}^{\beta}$ and $\overline{U}_{\;1}^{\alpha}$.
As this is a separate operation from the calculation of $\sqrt{\lambda_{n}}$
(Eq. \ref{eq:NTOHEinNGWFs}), we can first solve for a number of
eigenvalues $n$, where $n\le N_{N}$, and subsequently define a number
$m$, where $(m\le n$) of these where the associated singular vectors
are of interest. This can reduce the computational effort compared
to directly running an SVD algorithm given that many excitations will
be dominated by just NTO$1$ or a small number of NTO$n$ pairs.

\subsection{Interpretation of the transition origins of NTO$n$ electron-hole
pairs}

After obtaining the NTO-based transition vector element to any NTO$n$,
$n\le N_{N}$, next we want to interpret the transition origins of
each NTO$n$ electron-hole pair from the projection to a standard
orbital set. In the present work, we will in some places employ 
natural bond orbitals (NBOs)\cite{LC12} as the standard orbital set.
By such a projection
the composition of NTO$n$-H and NTO$n$-E can be quantitatively defined.
We generate NBOs by applying the NBO $5.0$\cite{GB11} program as a
postprocessing step after the ONETEP calculation. A relatively large
atomic orbital (AO) basis set is used in ONETEP for generating NBOs,
so that its quality is comparable to that of standard in-situ optimised NGWFs. 
The electron spillage is thus reduced to a low level, typically $0.006\%$
in the systems studied here, while the HOMO energy is typically within
$0.03$ eV of its value for NGWFs.

Note that the transition vector, $\mathbf{X}_{I}$,
can be based on different sets of reference orbitals in different
calculations. Rewriting the transition vector using the unique NTO
basis removes this reference dependence. Projection of NTOs onto a
common orbital set such as NBOs then provides a unification of interpretative
orbital set for these NTOs. For example, excitations in two systems
of different environments and/or geometries can be compared and the 
precise contribution of an orbital of interest
to an excitation can be computed.  We may be interested in the hole orbital
involved in a $^{1}\pi\pi^{\ast}$ excitation of a molecule, and can
compute its quantitative involvement in an excitation of the same
molecule embedded in a chemical/physical environment with or without
a covalent bond formed.

Two NTOs can be compared using the root mean square deviation, $\sigma$,
of the projection coefficients with respect to a common standard orbital
set\cite{LCGH12} such as the NBOs used in this work.
The similarity of two NTOs can also be quantified by 
the magnitude of projection from one to the other.
The projection sign can be arbitrary, as switching the overall phases
of both NTO$n$-H and NTO$n$-E orbitals at the same time will result
in them remaining a singular vector pair in Eq. \ref{eq:SVDinNGWFs}.
We will use the latter method for comparing two NTO$n$-H(E) orbitals 
in the present work, while the projection to NBOs is only employed to help 
interpret the transition origins of NTOs.

The transition origins of each excitation can be seen as its fingerprint, 
helping to identify excitations of similar character. The adiabatic potential 
energy surfaces of electronic states can then be `reconnected'
on the basis of their character.  If an electronic state 
of specific character changes its adiabatic energetic ordering from one 
geometry to the next, one would expect that at least one conical intersection 
(CX), key to many photochemical and photophysical processes
\cite{BRS94,Y01,DY12,MBL00,LM07,CB03,LCM08,RC98,PD06,KT02,CO95}, 
of electronic states has been passed between the two geometries.  A
by-product of analysing transition origins of a series of geometries along
some coordinate(s) is, therefore, that one observes the influence of a CX (hyper-)point
and locates the point where its projection onto the chosen path lies.
Note that the current method is not by
itself intended for locating the precise geometry of CX points, as to
do that would require searching through the full internal coordinate space.
In many well-defined cases, an appropriate series of geometries for analysing 
transition origins may be obtained by (constrained) geometry optimisation within the
ground state or a chosen excited state\cite{FA02}, by molecular 
dynamics for the ground state or excited states\cite{TTR07}, via transition state search
\cite{GPFKA03} etc. Overall, the method can be highly illustrative
as to how a photo-absorption driven process/reaction might proceed.

We also note that due to the approximate nature of practical TDDFT functionals,
the behaviour of TDDFT excitation energies around CXs may be incorrect\cite{LKQM06}, 
e.g. they may have overly rapid energy variation.  TDDFT can also 
give the wrong dimension of the branching plane\cite{LKQM06}. In other words, 
the PESs along some branching plane coordinates can 
be qualitatively wrong as the degeneracy is not lifted.  While these issues of 
TDDFT on the study of CXs are still highly debated, TDDFT has, 
however, been able to offer a reasonable description of oxirane 
photochemistry\cite{TT08,CD07} which will be our focus in the present work.
It has also been shown that TDA as implemented in the current work has some 
advantage over full TDDFT in the description of regions around intersections
\cite{HW14,CD07}.
It is known that the use of the TDA in TDDFT avoids the problems associated
with singlet and triplet instabilities\cite{CG00}. The former can
occur for systems which have strong static correlation, signified by the vanishing
HOMO-LUMO gap, whereas the latter can occur for systems which have an unstable
closed-shell ground state, reflected by a negative triplet excitation
energy within the TDDFT/TDA.

Overall, since locating CXs is not the 
point of the present method, the ability to describe the immediate vicinity of 
a CX is not critical.  As long as the influence of a CX on the excitation 
character of geometries reasonably distant from the CX itself in configuration 
space is qualitatively correct, then the method outlined will be useful.
Note also that the current QNTO method is not limited to use 
with TDDFT; any electronic structure methods that provide a decomposition
of excitations into single transitions of electrons can be combined with
this approach, and TDDFT is particularly useful for the study of very large molecular systems.

\begin{figure}[h]
\raggedright 

\begin{centering}
\includegraphics[natwidth=10.0cm,natheight=6.0cm,scale=0.82]{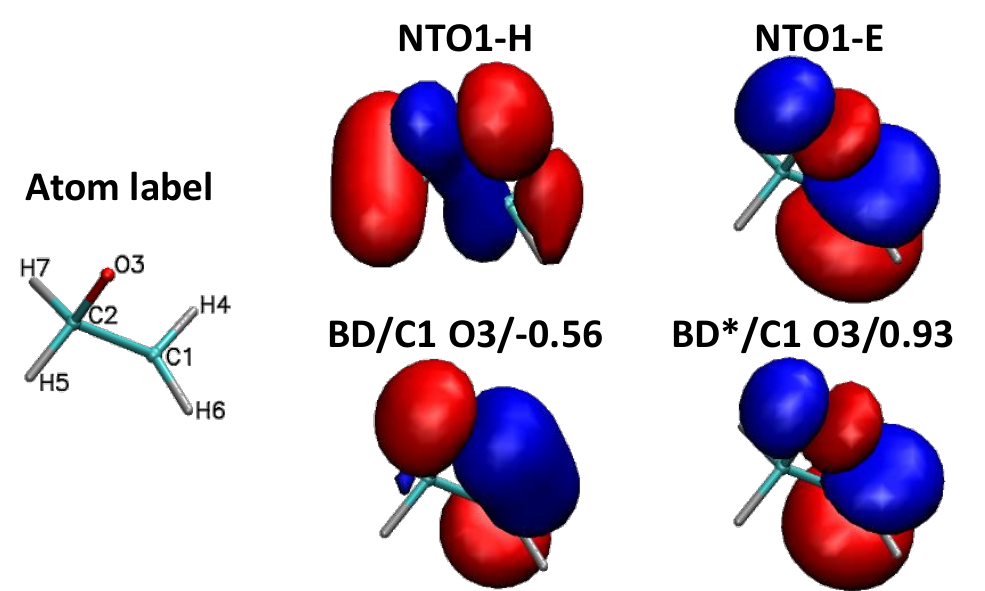}
\par\end{centering}

\caption{The NTO$1$-H and NTO$1$-E orbitals for the $2$nd excitation of
an oxirane molecule with CCO=$100\degree$. The $1$st dominant NBO
for each NTO$1$-H(E) and the projection coefficient is shown below. Isovalue=$0.05$ e/bohr$^3$. The NBO labels specify bonding
(BD) and anti-bonding (BD$^{\ast}$) and denote which pairs of atoms
(eg C1 and O3) are involved in a given bond.}

\label{fig:NTOexample} 
\end{figure}

As an example, we can examine the $2$nd excitation of an oxirane molecule
using TDDFT/QNTO. The geometry of the oxirane molecule, with CCO angle=$100\degree$,
is taken from a scan along the CCO angle which will be discussed in
greater detail in Sec. IV. There are $9$ valence (occupied) KS-orbitals
in the calculation, and therefore the HOMO(LUMO) is indexed $9$($10$)
etc. The transition density operator expressed in the KS-orbital basis
results from summation over several pairs of orbitals:
\begin{equation}
\begin{split}\hat{\rho}_{I=2}^{\{1\}}= & |\psi_{10}\rangle0.85\langle\psi_{8}|-|\psi_{10}\rangle0.35\langle\psi_{5}|-|\psi_{10}\rangle0.32\langle\psi_{6}|-\\
 & |\psi_{19}\rangle0.10\langle\psi_{9}|-|\psi_{14}\rangle0.09\langle\psi_{8}|+|\psi_{11}\rangle0.09\langle\psi_{8}|+\ldots
\end{split}
\end{equation}
By contrast, in the NTO basis it is mostly accounted for by a single
NTO pair: 
\begin{equation}
\begin{split}\hat{\rho}_{I=2}^{\{1\}}= & |\psi_{1}^{\{Nc\}}\rangle0.98\langle\psi_{1}^{\{Nv\}}|+|\psi_{2}^{\{Nc\}}\rangle0.13\langle\psi_{2}^{\{Nv\}}|+\\
 & |\psi_{3}^{\{Nc\}}\rangle0.09\langle\psi_{3}^{\{Nv\}}|+|\psi_{4}^{\{Nc\}}\rangle0.07\langle\psi_{4}^{\{Nv\}}|+\ldots
\end{split}
\label{Tran_vec_NTO}
\end{equation}
The transition density operator is dominated by the NTO$1$ transition,
which accounts for about $97\%$ of the total transition density,
whereas in the KS-orbital basis the largest component, $|\psi_{10}\rangle0.85\langle\psi_{8}|$,
only accounts for about $72\%$ of the density. In Fig. \ref{fig:NTOexample}
we plot the NTO$1$-H and NTO$1$-E orbitals as well as the most dominant
NBOs for each.

\section{Results and Discussion}

In this section, we discuss two applications of QNTO. In subsection
A, we focus on tracing the PESs associated with the ring opening
process of oxirane. In subsection B, we use QNTO to locate and compare
similar excitations between trans-2,3-diphenyloxirane and those
of oxirane.
As discussed in Sec. III B, it is interesting to see how environmental factors can
affect the transition origins of excitations of a molecule, which in turn may modify
its reaction pathways.  The study of trans-2,3-diphenyloxirane thus serves as
an excellent example of the use of QNTO to identify specific excitations of the 
oxirane molecule embedded in a different environment.

\begin{figure}
\raggedright 

\begin{centering}
\includegraphics[natwidth=33.867cm,natheight=19.05cm,scale=0.26]{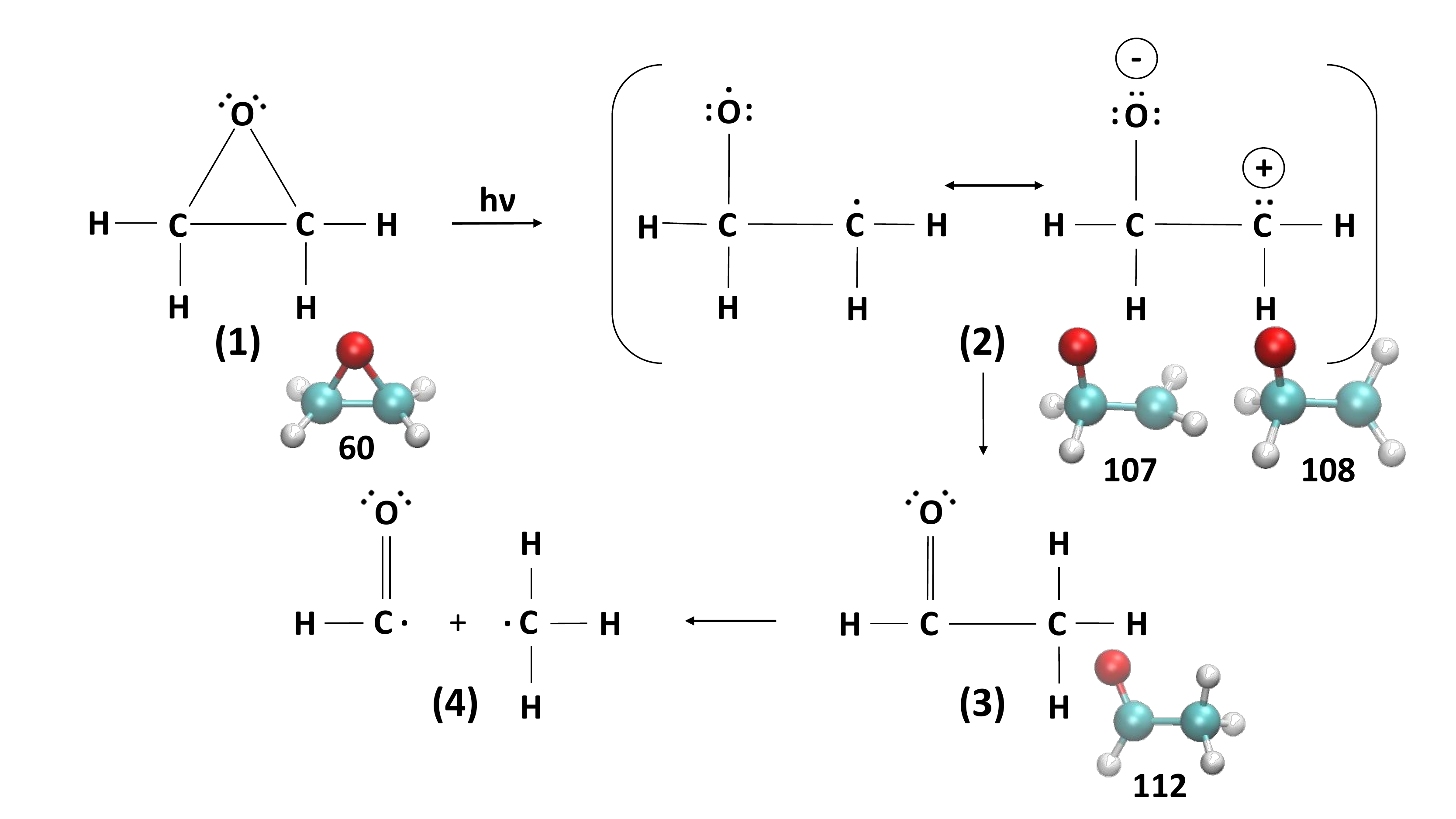}

\par\end{centering}

\caption{Gomer-Noyes mechanism. The underlying snapshots are from the CCO angle
scans discussed in the text.}

\label{fig:GN_pic} 
\end{figure}

\begin{figure}[h]
\raggedright 

\begin{centering}
\includegraphics[natwidth=25.4cm,natheight=16.0cm,scale=0.36]{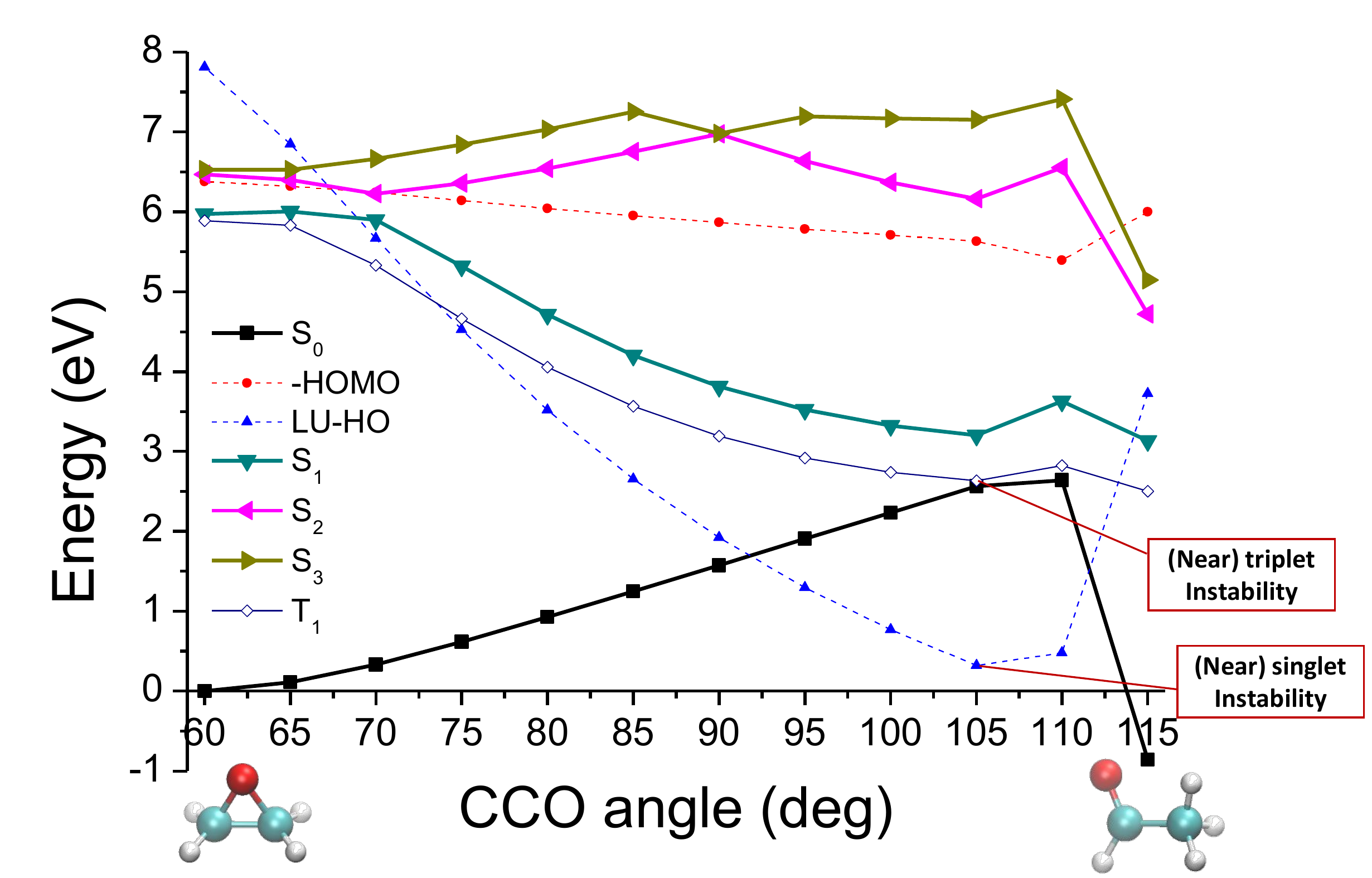}

\par\end{centering}

\caption{Constrained optimisation scan for different CCO angles followed by
TDLDA/TDA calculation. All internal coordinates during the scan are
optimised except for the CCO bond angle that is fixed in each step.}

\label{fig:CCOscan} 
\end{figure}

In all calculations a cut-off Coulomb approach\cite{HDHS11} has been
employed to avoid the influence of periodic images from neighbouring
simulation cells. A minimal set of $1$ NGWF for H atom and $4$ NGWFs
for C and O atoms is used. All calculations are well-converged
with respect to the variational parameters of the localisation radii:
suitable values are $10$ bohr for valence NGWFs and $16$ bohr for
conduction NGWFs. The kinetic energy cutoff for the grid spacing is
well-converged at $1000$ eV. Norm-conserving pseudopotentials are
used to replace core electrons.
For DFT and TDDFT calculations we make use of the (A)LDA functional 
in the Perdew-Zunger parameterisation\cite{PZ81} throughout.
While checking the accuracy of results using other higher level method 
may be desirable, we note that DFT/TDDFT calculations with the standard 
(semi-)local functionals LDA and PBE\cite{PBE96} are capable of 
describing the photochemically relevant PESs of ring-opening reactions 
in qualitative agreement with high quality quantum Monte Carlo 
calculations\cite{TT08,CD07}.
In addition, even though the use of Hartree-Fock exchange hybridised 
functionals such as PBE0\cite{AB99} or asymptotically corrected functionals 
such as LB94\cite{LB94} may improve the DFT/TDDFT results,
the results can deteriorate to be worse than that of pure PBE 
for structures near bond breaking region at large CCO angle where 
correlation effects become highly significant\cite{TT08}.
Although further examinations of functionals such as range-separated 
hybrid ones that have been used for similar molecular systems
\cite{CH08,WPS09,SG11,LCGH11} may still be desirable, 
the (A)LDA is considered to be sufficient for the purpose of the
present work.

\begin{sidewaysfigure}[H]
\includegraphics[natwidth=12.788cm,natheight=4.048cm,scale=1.85]{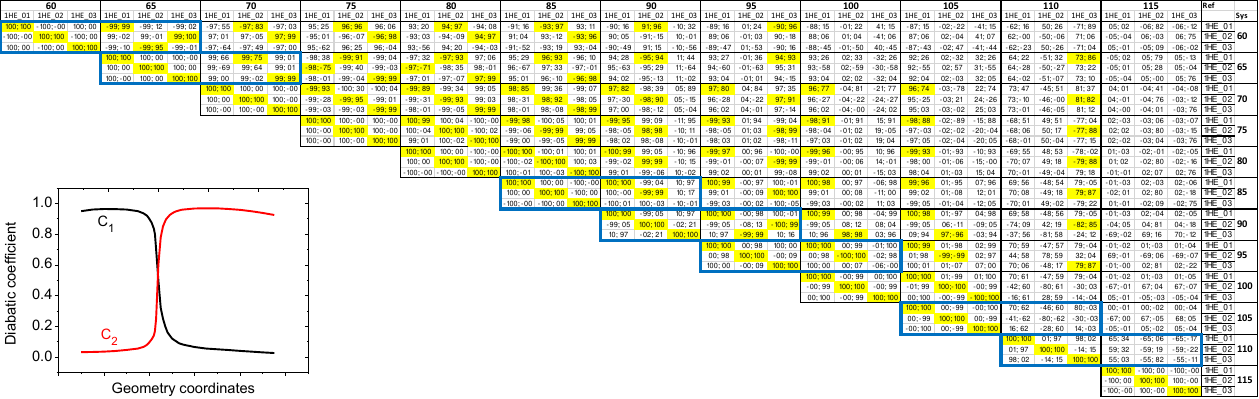}
\caption{Projection results for different pairs of NTO$1$-H(E) orbitals. Each
row/column represents one of the first three excitations at each geometry
at $5\degree$ intervals in a scan in the range CCO=$60-115\degree$.
Each pair of CCO angles form a block in the map, which is divided
into $9$ cells for different pairing of excitations. The first and
second number in a cell is the projection result (\%) for NTO$1$-H
and NTO$1$-E, respectively. A cell with yellow background indicates
that both numbers are $\ge1/\sqrt{2}$. The blocks marked by blue
lines are where some character changes of excitations are observed
one step off the diagonal blocks due to, for example, passing a CX.
The small figure shows an example of adiabatic state behaviour near
a CX\cite{AR97}. C$_{n}$: (Diabatic) configuration coefficient
for the two adiabatic states.}

\label{fig:60-115map} 
\end{sidewaysfigure}

\subsection{Photoinduced ring opening of oxirane}

\begin{table}
\caption{The transition vector component for NTO$1$ ($\sqrt{\lambda_{1}}$)
of the first eight excitations along the oxirane CCO scan (Fig. \ref{fig:CCOscan}).}
\footnotesize
\begin{ruledtabular}
\begin{tabular}{ccccccccc}
\bf CCO &     \multicolumn{ 8}{c}{\bf Excitation number}     \\
\bf angle &    {\bf 1} &    {\bf 2} &    {\bf 3} &    {\bf 4} &    {\bf 5} &    {\bf 6} &    {\bf 7} &    {\bf 8} \\
\hline
       \bf 60 &     0.9997 &     0.9999 &     0.9997 &     0.9981 &     0.9997 &     0.9985 &     0.9992 &     0.9961 \\
\hline
       \bf 65 &     0.9997 &     0.9997 &     0.9999 &     0.9980 &     0.9994 &     0.9986 &     0.9976 &     0.9980 \\
\hline
       \bf 70 &     0.9995 &     0.9997 &     0.9999 &     0.9976 &     0.9992 &     0.9916 &     0.9900 &     0.9983 \\
\hline
       \bf 75 &     0.9997 &     0.9998 &     0.9999 &     0.9964 &     0.9989 &     0.8737 &     0.8832 &     0.9947 \\
\hline
       \bf 80 &     0.9998 &     0.9999 &     0.9998 &     0.9921 &     0.9420 &     0.9874 &     0.9874 &     0.7348 \\
\hline
       \bf 85 &     0.9999 &     0.9998 &     0.9992 &     0.8237 &     0.7862 &     0.9972 &     0.9978 &     0.9560 \\
\hline
       \bf 90 &     0.9999 &     0.8798 &     0.8477 &     0.9971 &     0.9977 &     0.9926 &     0.9988 &     0.9789 \\
\hline
       \bf 95 &     0.9999 &     0.9800 &     0.9967 &     0.9963 &     0.9997 &     0.9954 &     0.9809 &     0.9996 \\
\hline
      \bf 100 &     0.9999 &     0.9843 &     0.9994 &     0.9996 &     0.9997 &     0.9859 &     0.9733 &     0.9877 \\
\hline
      \bf 105 &     0.9999 &     0.9868 &     0.9996 &     0.9997 &     0.9997 &     0.7387 &     0.7487 &     0.9853 \\
\hline
      \bf 110 &     0.9989 &     0.9503 &     0.9810 &     0.9067 &     0.9952 &     0.8161 &     0.7883 &     0.8233 \\
\hline
      \bf 115 &     0.9997 &     0.9959 &     0.9992 &     0.9993 &     0.9983 &     0.9909 &     0.9965 &     0.9877 \\
\end{tabular}
\end{ruledtabular}
\label{tab:60-115SEC}
\end{table}

\begin{figure*}
\includegraphics[natwidth=8.021cm,natheight=2.562cm,scale=1.85]{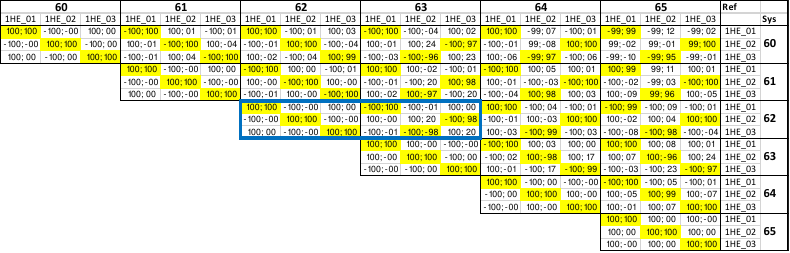}
\caption{Projection results for different pairs of NTO$1$-H(E) orbitals. Each
orbital comes from the first three excitations of a CCO=$60-65\degree$
scan. An abrupt switch of transition origins betweeen the second and third
excited state can be seen to occur between $62\degree$ and $63\degree$.}

\label{fig:60-65map} 
\end{figure*}

\begin{figure}
\raggedright 

\begin{centering}
\includegraphics[natwidth=25.4cm,natheight=19.05cm,scale=0.37]{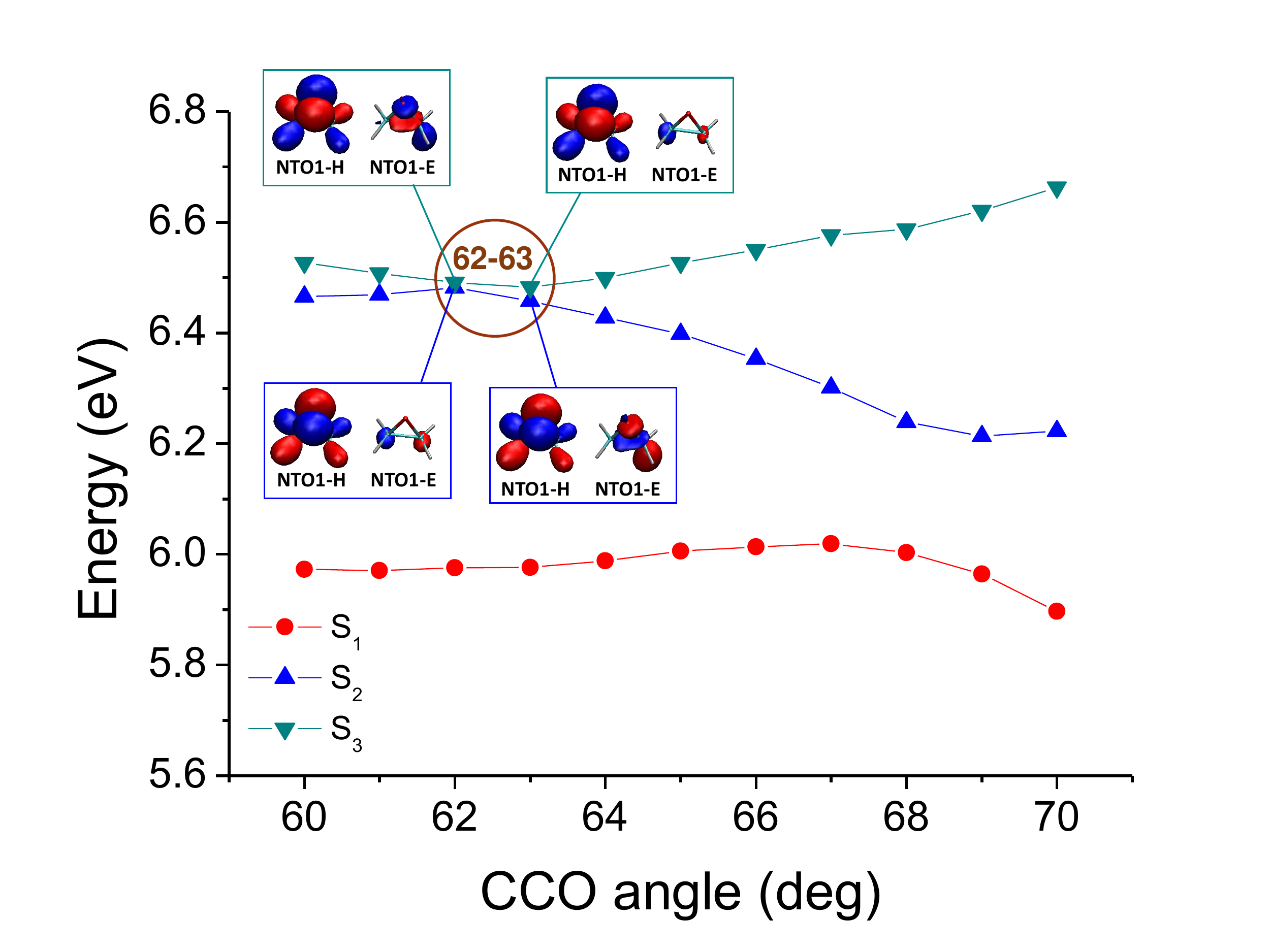}
\par\end{centering}

\caption{The corresponding potential energy curves for the map shown in Fig. \ref{fig:60-65map}.
The NTO$1$-H(E) orbitals at $62\degree$ and $63\degree$ have also
been plotted, with isovalue=$0.05$ e/bohr$^3$. While the NTO$1$-H remains the
same for the $2$nd and $3$rd excitation, the NTO$1$-E orbitals
can be seen to have switched their features between the $2$nd and
$3$rd excitation at $62\degree$ and $63\degree$, namely, the NTO$1$-E
character of the $2$nd($3$rd) excitation at $62\degree$ corresponds
to that of the $3$rd($2$nd) excitation at $63\degree$.}

\label{fig:60-65PES} 
\end{figure}

\begin{figure*}
\includegraphics[natwidth=8.021cm,natheight=2.562cm,scale=1.85]{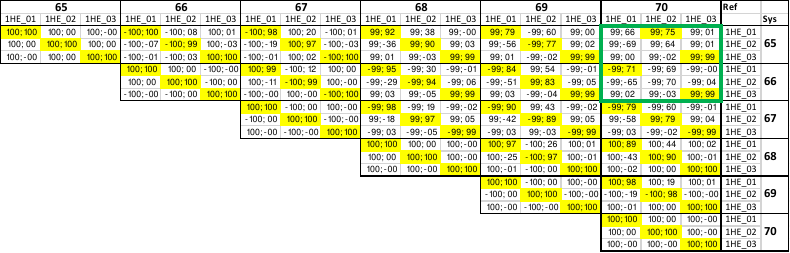}
\caption{Projection results for different pairs of NTO$1$-H(E) orbitals. Each
orbital comes from the first three excitations of a CCO=$65-70\degree$
scan. The green block indicates that, using the NTO$1$-H and NTO$1$-E
of the first two excitations at CCO=$70\degree$ as the reference
set of orbitals, a switch of the dominating electron orbitals is observed
between CCO=$65\degree-66\degree$. For more details see the text.}

\label{fig:65-70map} 
\end{figure*}

\begin{figure}
\raggedright 

\begin{centering}
\includegraphics[natwidth=25.4cm,natheight=19.05cm,scale=0.37]{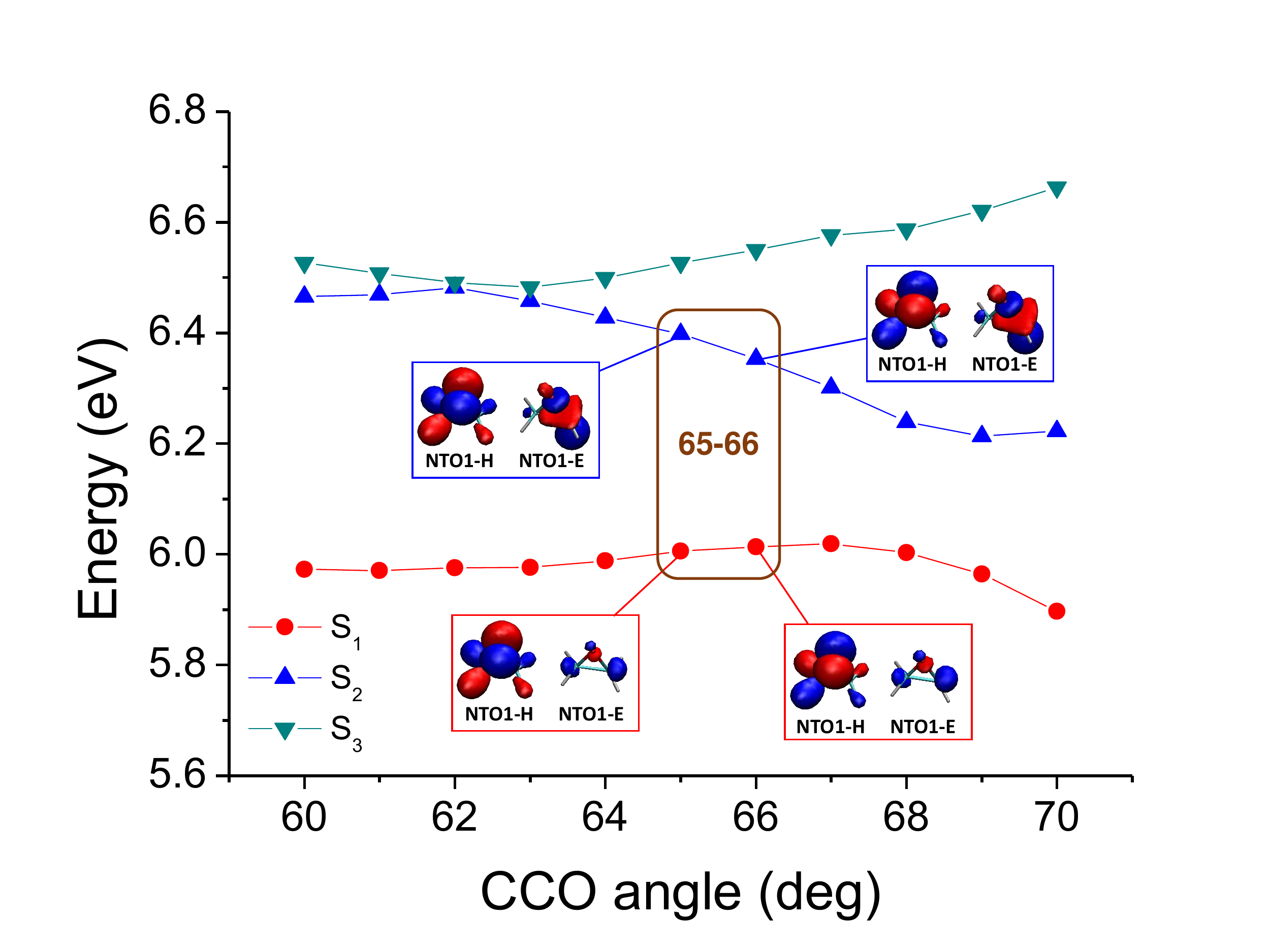}
\par\end{centering}

\caption{The corresponding potential energy curves for the map shown in Fig. \ref{fig:65-70map}.
Unlike the case in Fig. \ref{fig:60-65map}, the switch of dominant
NTO character cannot be seen from the plot of NTO$1$-H and NTO$1$-E,
reflected also by the map (Fig. \ref{fig:65-70map}) where the projection
results among $65\degree$, $66\degree$, and $67\degree$ cannot be seen to reveal a switch
due to a smoother transition of dominant NTO$1$ character.}

\label{fig:65-70PES} 
\end{figure}

\begin{figure}
\raggedright 

\begin{centering}
\includegraphics[natwidth=25.4cm,natheight=19.05cm,scale=0.37]{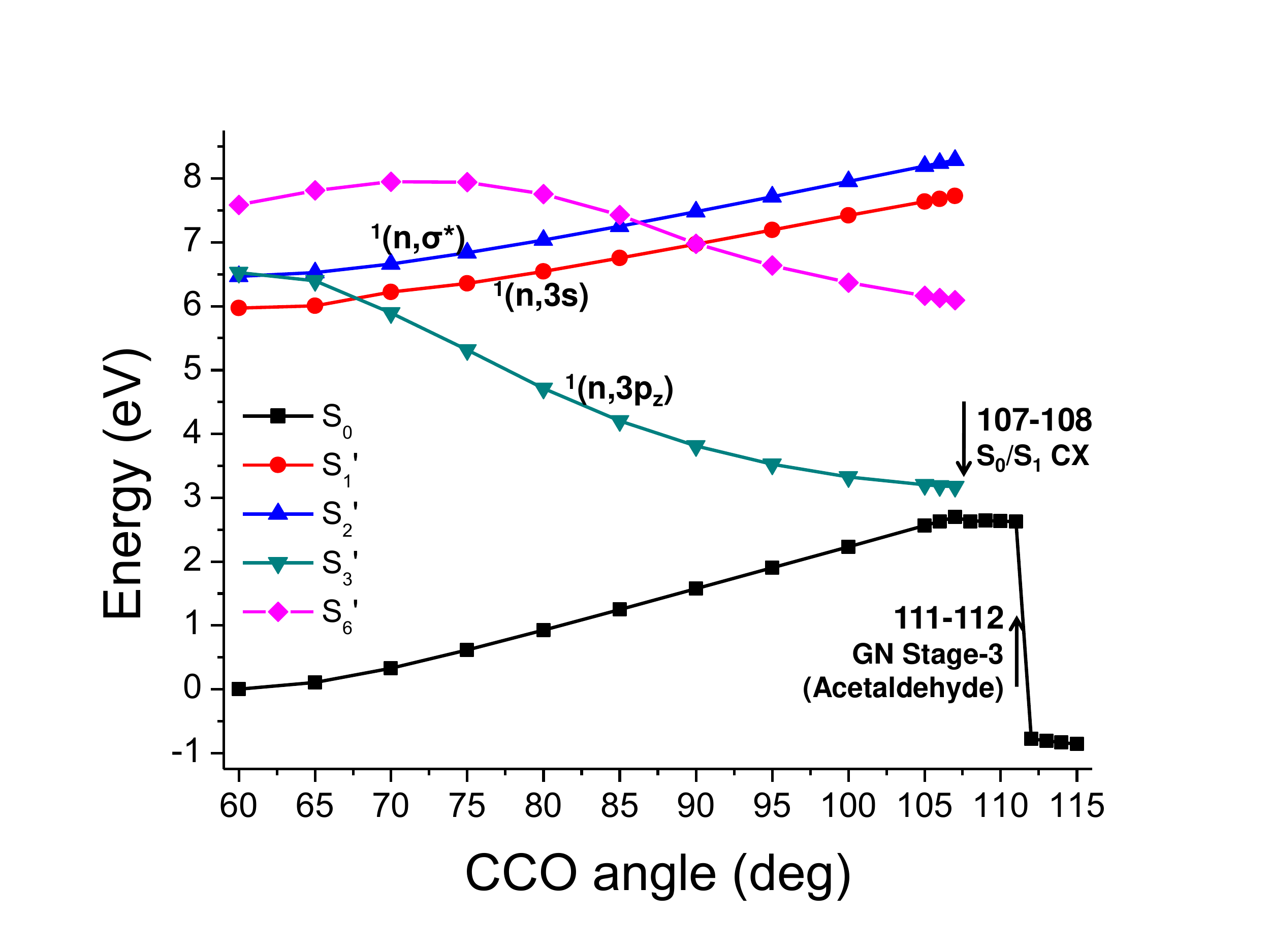}
\par\end{centering}

\caption{The $1$st, $2$nd, $3$rd, and $6$th singlet excitation potential
energy curve (ranked at CCO=$60\degree$) that is re-connected based
on the similarity of transition origins.}

\label{fig:reconPES} 
\end{figure}

\begin{figure*}
\raggedright 

\begin{centering}
\includegraphics[natwidth=20.0cm,natheight=6.0cm,scale=0.9]{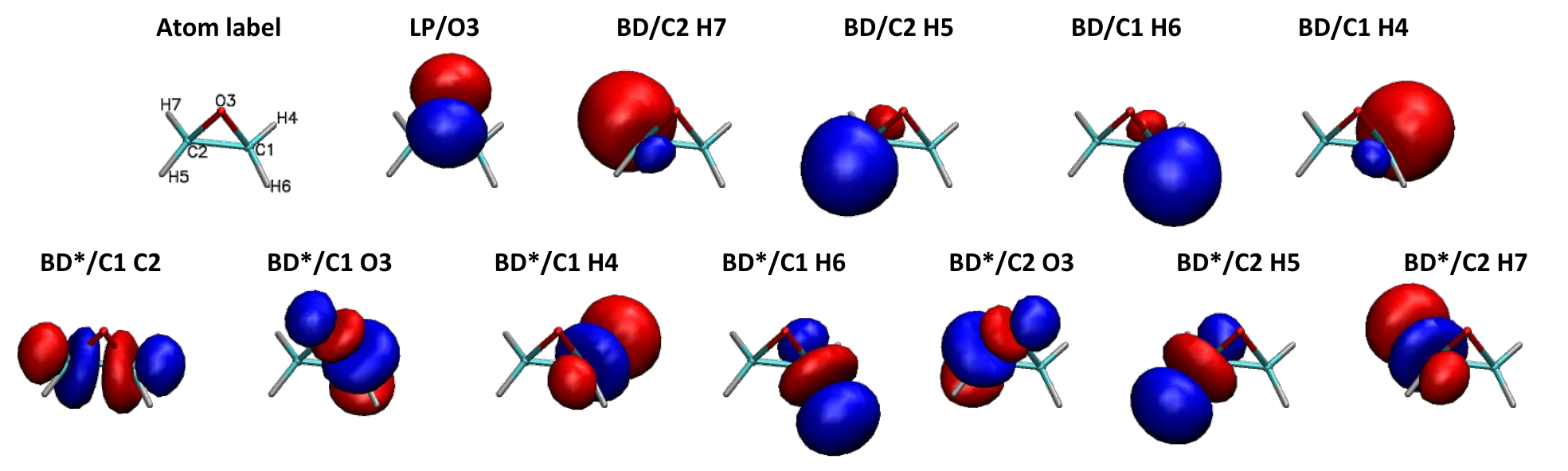}
\par\end{centering}

\caption{NBOs corresponding to Table \ref{tab:QNTOinfo60}. The lone pair
on the oxygen (LP/O$3$) accounts for a large proportion of the NTO$1$-H
for all three excitations. The $1$st excitation is represented by
NBOs denoted BD$^{\ast}$/C$2$ H$7$, BD$^{\ast}$/C$1$ H$4$, BD$^{\ast}$/C$1$
H$6$, and BD$^{\ast}$/C$2$ H$5$, whereas the $2$nd excitation
is represented by the same set of NBOs each with a reduced contribution,
plus the more important BD$^{\ast}$/C$1$ H$2$. The $3$rd excitation
is represented by BD$^{\ast}$/C$1$ O$3$ and BD$^{\ast}$/C$2$
O$3$.}

\label{fig:NBOs} 
\end{figure*}

The Gomer-Noyes mechanism, as depicted in Fig. \ref{fig:GN_pic}, was
postulated and experimentally confirmed\cite{GN50,KIIT73} as a way to
explain the photoinduced ring opening process of the oxirane molecule.
More discussions about this textbook molecule of stereochemistry can be
found, for example, in Ref. \onlinecite{MB90}.
In particular, the specific electronic states involved in the Gomer-Noyes
mechanism have been examined in some detail through TDDFT excited state
molecular dynamics (MD) simulations validated by high-quality quantum
Monte Carlo calculations\cite{TT08,CD07}, and several states have been
found to be involved in mediating the ring-opening reaction.

Here, we perform a constrained ground state optimisation scan along the CCO bond
angle of oxirane. We relax all internal coordinates except for a fixed
CCO bond angle, which is set to values in the range $65\degree$ to
$110\degree$. For each oxirane geometry in the scan, a TDDFT calculation
(vertical excitations) is performed followed by QNTO analysis to obtain 
the transition origins of the first few low-lying excitations.

The potential energy curves of the low-lying singlet and triplet states
are plotted in Fig. \ref{fig:CCOscan}. For reference we also show
the negative HOMO energy and the value of the HOMO-LUMO gap at each
angle. A caveat is that the calculation quality
would deteriorate for those excitations of energies higher than the
negative HOMO energy. As this tends to be underestimated by the TDDFT/LDA\cite{CJCS98},
one may obtain fewer bound states than are present in the real system.
In addition, as mentioned in Sec. III. B, a vanishing HOMO-LUMO gap and triplet 
excitation energy signifies a (nearly) singlet and triplet
instability, respectively, whereas employment of TDA helps circumvent these 
problems and may be recommendable when using generic approximate functionals
in TDDFT calculations.

We note that from the excitation energies alone there is no clear
indication of where crossings, indicating nearby CX points, have been passed
while progressing along these curves. This is where the utility of QNTO analysis
can be seen, as discussed in Sec. III B.
If the transition density operator for these systems is dominated
by the NTO$1$ component (Eq. \ref{Tran_vec_NTO}), the character of
each excitation can be represented by the transition origins of just NTO$1$.
The transition vector component for NTO$1$, $\sqrt{\lambda_{1}}$,
along the scan are listed in Table \ref{tab:60-115SEC}. As can
be seen, in most cases $\sqrt{\lambda_{1}}$ is higher than $0.99$.
For the first three excitations, the smallest $\sqrt{\lambda_{1}}$,
$0.8477$, is from the $3$rd excitation of the oxirane geometry at
CCO=$90\degree$, which still captures about $72\%$ of the density
contribution. Thus, all the excitations on the oxirane geometries
from the scan are dominated by the NTO$1$ transition, and when a
switch of the NTO$1$ transition origins occurs, it shows that the
excitation changes its character.

To put this in a form suitable for identifying electronic states of
similar character section by section, we can construct
a map (Fig. \ref{fig:60-115map}) showing the projections of the
NTO$1$ pairs for each of the first three excitations ($S_{1}$, $S_{2}$,
$S_{3}$) at each geometry onto their equivalents at different geometries,
forming a $3\times3$ block for each pair of geometries.
We refer to the geometry listed down the row of the map as the
`System' (Sys) and compare it to a `Reference' (Ref) geometry along the
columns. To obtain a meaningful projection while accounting for the geometry
change, the NTO$1$-H(E) orbitals are translated from the Sys-geometry
to the Ref-geometry and renormalised before being used for projection. 
The optimised atom-centred NGWFs $|\phi_{\alpha;\textrm{s}}\rangle$
from the calculation in the Sys-geometry are translated to the corresponding
atom centres in the Ref-geometry to produce NGWFs $|\phi_{\alpha;\textrm{s}\rightarrow\textrm{r}}\rangle$.
The NTO$1$-H orbital (Eq. \ref{eq:NTOinNGWFs}) is then re-expressed
in terms of these using the original coefficients $\overline{V}_{\;1;\textrm{s}}^{\alpha}$
\begin{equation}
|\psi_{1;\textrm{s}\rightarrow\textrm{r}}^{\{Nv\}}\rangle=|\phi_{\alpha;\textrm{s}\rightarrow\textrm{r}}\rangle\overline{V}_{\;1;\textrm{s}}^{\alpha}\;.
\end{equation}
The translated orbital is then renormalised: 
\begin{equation}
|\psi_{1;\textrm{s}\rightarrow\textrm{r}}^{\prime\{Nv\}}\rangle=\frac{|\psi_{1;\textrm{s}\rightarrow\textrm{r}}^{\{Nv\}}\rangle}{\sqrt{\langle\psi_{1;\textrm{s}\rightarrow\textrm{r}}^{\{Nv\}}|\psi_{1;\textrm{s}\rightarrow\textrm{r}}^{\{Nv\}}\rangle}}\;,
\label{eq:renormal}
\end{equation}
and each $3\times3$ block of projections is calculated as
\[
\langle\psi_{I;\textrm{s}\rightarrow\textrm{r}}^{\prime\{Nv\}}|\psi_{J;\textrm{r}}^{\prime\{Nv\}}\rangle;\langle\psi_{I;\textrm{s}\rightarrow\textrm{r}}^{\prime\{Nc\}}|\psi_{J;\textrm{r}}^{\prime\{Nc\}}\rangle
\]
for $I=1,2,3$ and $J=1,2,3$.

\begin{table}
\caption{Decomposition of NTO$1$-H(E) into NBOs for the first three excitations
of oxirane at CCO=$60\degree$. $N$ indicates the excitation number.
Proj denotes the projection result. NBO type gives information on
the NBO feature. For each excitation, the first $6$ ranked NBOs of
lone pair (LB), bonding (BD), and anti-bonding (BD$^{\ast}$) types
are listed. The highly basis-dependent Rydberg-type NBOs (RY$^{\ast}$)
that accounts for most of these excitations are not listed. Note that
the projection magnitude of the $6$th-ranked NBO has gone down to
below $0.1$ for all cases. Total means the total density contribution
(square sum) of all the $6$ NBOs listed. These NBOs are plotted in
Fig. \ref{fig:NBOs}.}

\begin{ruledtabular}
\begin{tabular}{r|llr|llr}
\multicolumn{1}{c|}{} & \multicolumn{3}{c|}{NTO1-H} & \multicolumn{3}{c}{NTO1-E}\tabularnewline
\multicolumn{1}{r|}{N} & \multicolumn{2}{c}{NBO type} & Proj  & \multicolumn{2}{c}{NBO type} & Proj \tabularnewline
\hline 
\multicolumn{1}{c|}{1} & LP  & O3  & 0.894  & BD{*}  & C2 H7  & -0.168 \tabularnewline
\multicolumn{1}{c|}{} & BD  & C2 H7  & -0.215  & BD{*}  & C1 H4  & -0.167 \tabularnewline
\multicolumn{1}{c|}{} & BD  & C2 H5  & 0.213  & BD{*}  & C1 H6  & -0.165 \tabularnewline
\multicolumn{1}{c|}{} & BD  & C1 H6  & 0.199  & BD{*}  & C2 H5  & -0.164 \tabularnewline
\multicolumn{1}{c|}{} & BD  & C1 H4  & -0.198  & BD{*}  & C1 O3  & 0.055 \tabularnewline
\multicolumn{1}{c|}{} & BD{*}  & C2 H5  & 0.080  & BD{*}  & C2 O3  & -0.048 \tabularnewline
\multicolumn{1}{r|}{} & \multicolumn{2}{r}{Total} & 0.976  & \multicolumn{2}{r}{Total} & 0.116 \tabularnewline
\hline 
\multicolumn{1}{c|}{2} & LP  & O3  & -0.894  & BD{*}  & C1 C2  & -0.210 \tabularnewline
\multicolumn{1}{c|}{} & BD  & C2 H7  & 0.215  & BD{*}  & C1 H4  & 0.124 \tabularnewline
\multicolumn{1}{c|}{} & BD  & C2 H5  & -0.213  & BD{*}  & C2 H5  & -0.123 \tabularnewline
\multicolumn{1}{c|}{} & BD  & C1 H6  & -0.200  & BD{*}  & C1 H6  & 0.121 \tabularnewline
\multicolumn{1}{c|}{} & BD  & C1 H4  & 0.198  & BD{*}  & C2 H7  & -0.119 \tabularnewline
\multicolumn{1}{c|}{} & BD{*}  & C2 H5  & -0.080  & BD{*}  & C1 O3  & 0.019 \tabularnewline
\multicolumn{1}{r|}{} & \multicolumn{2}{r}{Total} & 0.976  & \multicolumn{2}{r}{Total} & 0.104 \tabularnewline
\hline 
\multicolumn{1}{c|}{3} & LP  & O3  & 0.890  & BD{*}  & C1 O3  & -0.306 \tabularnewline
\multicolumn{1}{c|}{} & BD  & C2 H7  & -0.219  & BD{*}  & C2 O3  & 0.270 \tabularnewline
\multicolumn{1}{c|}{} & BD  & C2 H5  & 0.218  & BD{*}  & C2 H7  & -0.077 \tabularnewline
\multicolumn{1}{c|}{} & BD  & C1 H6  & 0.203  & BD{*}  & C1 H4  & -0.076 \tabularnewline
\multicolumn{1}{c|}{} & BD  & C1 H4  & -0.202  & BD{*}  & C1 H6  & -0.067 \tabularnewline
\multicolumn{1}{c|}{} & BD{*}  & C2 H5  & 0.080  & BD{*}  & C2 H5  & -0.065 \tabularnewline
\multicolumn{1}{r|}{} & \multicolumn{2}{r}{Total} & 0.977  & \multicolumn{2}{r}{Total} & 0.187 \tabularnewline
\end{tabular}
\end{ruledtabular}
\label{tab:QNTOinfo60} 
\end{table}

In Fig. \ref{fig:60-115map} one can immediately see which cells
involve pairs of excitations with similar transition origins. We highlight
those that have both projections greater than $1/\sqrt{2}$, indicating
that the density contribution to one from the other is higher than
$50\%$. One then sees where there are switches of transition origins
of these adiabatic excited state curves, or changes of transition
character of all excited states, from one block to another. The former
case can indicate that there has been one or more CXs passed between
the two excitations. The latter would be caused by a character change
of the ground state itself. This can be due to the fact that a CX
between S$_{0}$ and S$_{1}$ has been passed, e.g. $105\degree$ and $110\degree$,
or that the molecular structures have become drastically different,
e.g. $105\degree$ and $65\degree$.

One can further subdivide the geometry scan to more accurately locate
the position along the CCO angle coordinate where the character of
the excitation changes.
We demonstrate two such examples of switches of excitation
which show subtly different behaviour. Fig. \ref{fig:60-65map} shows
an example where the switch of excitation character previously shown
from the coarser map to occur between $60\degree$ and $65\degree$
is narrowed down to occurring abruptly between $62\degree$ and $63\degree$.
The corresponding potential energy curves for the two states involved,
and the NTO$1$-H and NTO$1$-E orbitals before and after the switch,
are plotted in Fig. \ref{fig:60-65PES}, showing that the bands approach
very closely around this point.

In other cases, however, when a more closely spaced geometry map is
considered, the change of excitation character may be so smooth
that no abrupt character switch occurring between two
minimally-differentiated geometries can be found. An example is shown in Fig. \ref{fig:65-70map}.
The two adiabatic excited state curves as plotted in Fig. \ref{fig:65-70PES}
also do not come closer than $0.2$ eV to each other. However, as
the transition characters have switched for the two adiabatic excited
states curves between CCO=$65\degree-70\degree$, there must be a
CX located nearby, i.e. on other (internal) coordinates, even if it 
does not occur exactly along the scanned coordinate. 
Thus, without analysing the electronic character, only from the plot
of the energy, one would not necessarily know that molecular motions in this
region of the scanned coordinate would be near a CX, and that an electronic
state change from the upper surface to the lower surface could occur.
However, note that from considering the overall shape of the energies
in Fig. \ref{fig:CCOscan}, it is not difficult to envisage that a
crossing could have occurred between S$_1$ and S$_2$.
For further analysis, the NTO$1$-H(E) orbitals of excitations
from a geometry somewhat further away from those being considered
can be used as a reference set, similar to a diabatic configurations
set used for locating CXs\cite{AR97}. For the map shown in Fig. \ref{fig:65-70map},
the NTO$1$-H and NTO$1$-E orbitals of the first two excitations
at CCO=$70\degree$ can provide an appropriate reference set. We thus
project the NTO1-H and NTO1-E orbitals at varying geometries onto
these reference orbitals, giving: 
\begin{equation}
\begin{split} & |\psi_{1;I=1}^{\{Nc\}65\degree}\rangle=0.66|\psi_{1;I=1}^{\{Nc\}70\degree}\rangle+0.75|\psi_{1;I=2}^{\{Nc\}70\degree}\rangle+\ldots\\
 & |\psi_{1;I=2}^{\{Nc\}65\degree}\rangle=-0.69|\psi_{1;I=1}^{\{Nc\}70\degree}\rangle+0.64|\psi_{1;I=2}^{\{Nc\}70\degree}\rangle+\ldots
\end{split}
\label{eq:65in70}
\end{equation}
\begin{equation}
\begin{split} & |\psi_{1;I=1}^{\{Nc\}66\degree}\rangle=0.71|\psi_{1;I=1}^{\{Nc\}70\degree}\rangle+0.69|\psi_{1;I=2}^{\{Nc\}70\degree}\rangle+\ldots\\
 & |\psi_{1;I=2}^{\{Nc\}66\degree}\rangle=-0.65|\psi_{1;I=1}^{\{Nc\}70\degree}\rangle+0.70|\psi_{1;I=2}^{\{Nc\}70\degree}\rangle+\ldots
\end{split}
\;.\label{eq:66in70}
\end{equation}
This shows that the two reference electron orbitals from CCO=$70\degree$
switch domination between CCO=$65\degree$ and $66\degree$, which
could not have been determined from Fig. \ref{fig:65-70PES} alone.
Thus, a better strategy without overlooking a character switch between
two curves would be to employ a wide coordinate window within 
which all the cross-projection data between each pair of geometries 
are checked.

We note that the exact point where the domination switches occur can
be somewhat affected by the choice of reference orbitals used, although
this does not in general prevent us from tracing down the CX to a
small region in the scanned coordinate. In the language of diabatic
configurations\cite{AR97}, the situation can be analysed by considering
the matrix elements of a Hamiltonian $\hat{H}$ acting on a reduced
subspace consisting of just two such configurations, denoted as $|\Phi_{1}\rangle$
and $|\Phi_{2}\rangle$. The switch of domination of two diabatic
configurations as a given geometry path is traversed indicates that
the surface $\textrm{H}_{11}-\textrm{H}_{22}=0$ has been crossed,
though the exact shape of the surface of $\textrm{H}_{11}-\textrm{H}_{22}=0$
will be affected by the reference diabatic configurations used.

Evidence of a CX between S$_{1}$ and S$_{0}$, on the other hand, would be
reflected in the map by the fact that a whole column of excitations
would change transition character at the same time,
because the ground state itself on which excitations are based has
changed character. We are able to locate two such points in Fig. \ref{fig:60-115map}:
between $107\degree$ and $108\degree$, and between $111\degree$
and $112\degree$, both related to the change of the ground state character
expected from the Gomer-Noyes mechanism.

After locating the regions where two adiabatic excited state curves
switch their characters, we can plot the adiabatic singlet states
again, with the curves for the various excitations `re-connected'
based on their transition origins, as shown in Fig. \ref{fig:reconPES}.
State labels of $^{1}$(n,$3p_{z}$), $^{1}$(n,$3s$) and $^{1}$(n,$\sigma^{\ast}$)
are based on the symmetry at the CCO=$60\degree$ geometry and propagated
to other geometries based on the connection of the states.

It then becomes clear that population of the S$'_{3}$ state at CCO=$60\degree$
can directly lead to a crossing between S$_{0}$ and S$_{1}$. The
energy decreases monotonically along with the widening CCO angle, which
will result in a rapid ring opening reaction. This is consistent with
the previous excited states MD study\cite{TT08} which found that
the population of the $^{1}$(n,$3p_{z}$) state would facilitate
the occurrence of ring opening process. If the $^{1}$(n,$3$s) state
is populated instead, it has to overcome an energy barrier to a widening
of CCO angle before meeting a crossing with the $^{1}$(n,$3p_{z}$)
state which can then lead to the ring opening of oxirane. Thus, this
chemical process is more difficult and would produce a lower quantum
yield. We identify that the ground state is re-populated from the
$^{1}$(n,$3p_{z}$) state after oxirane passing the critical point
between $107\degree$ and $108\degree$.

Meanwhile, passing the critical point between $111\degree$ and $112\degree$
leads to another change of ground state character, which corresponds
to a proton transfer process producing acetaldehyde, the product in
the $3$rd stage of the Gomer-Noyes mechanism (Fig. \ref{fig:GN_pic}).
This is also consistent with the MD study\cite{TT08} finding that
the proton shift occurs after the oxirane has been de-excited to the
ground state. Under thermally elevated conditions this proton shift
will likely occur at lower CCO angles: here we address only the quasistatic
constrained geometries. A follow-up reaction breaking the C-C bond,
the final stage in Fig. \ref{fig:GN_pic}, would then be possible
while remaining in the ground state, given an appropriate combination
of internal motion\cite{TT08}.

To demonstrate the chemical composition of these transitions, a decomposition
of NTO$1$-H and NTO$1$-E into natural bond orbitals for the first
three excitations at CCO=$60\degree$ is shown in Table \ref{tab:QNTOinfo60}.
The AO basis used for generating NBOs is fairly large, producing $525$
NBOs, but these are mostly of Rydberg (RY$^{\ast}$) type. In practice,
the lone-pair (LP) and bonding (BD) NBOs describe the majority of
the NTO1-H excitation, which in all three cases is well-described
by just five main NBOs. For the NTO$1$-E orbitals of the three excitations,
however, a substantial contribution comes from RY$^{\ast}$
orbitals, which are highly dependent on the AO basis set used in generating
NBOs. The LP and BD NBOs for NTO1-H, and the most relevant BD$^{\ast}$
NBOs for NTO1-E are shown in Fig. \ref{fig:NBOs}.

\subsection{Trans-2,3-Diphenyloxirane}

\begin{figure}
\raggedright 

\begin{centering}
\includegraphics[natwidth=8.0cm,natheight=3.0cm,scale=0.85]{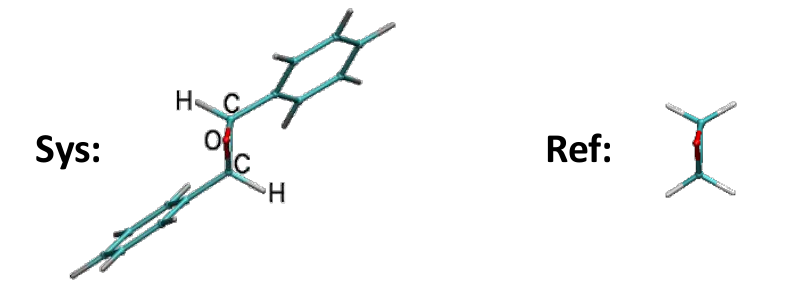}
\par\end{centering}

\caption{A plot of the system molecule, trans-2,3-diphenyloxirane, and the
reference oxirane molecule that is extracted from the trans-2,3-diphenyloxirane.
The five atoms labelled are the common (core) atoms.}

\label{fig:Mol_label} 
\end{figure}

\begin{figure*}
\raggedright 

\begin{centering}
\includegraphics[natwidth=40.0cm,natheight=20.0cm,scale=0.6]{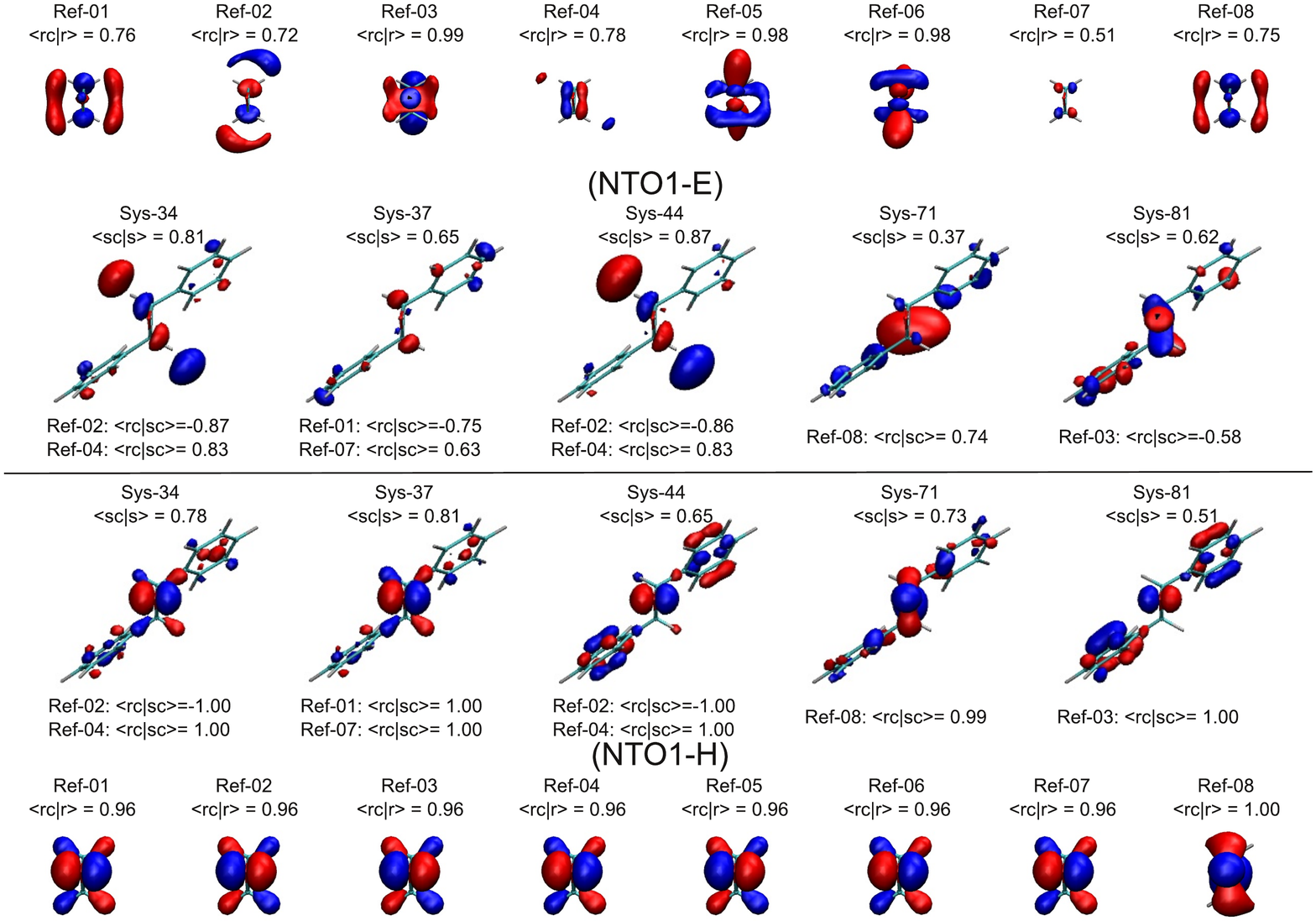} 
\par\end{centering}

\caption{Plots of the NTO$1$-H(E) orbitals for the first eight Ref-excitations
as well as several Sys-excitations that have a substantial overlap $|\langle\textrm{rc}|\textrm{sc}\rangle|$
to some Ref-excitations. Isovalue=$0.06$ e/bohr$^3$ for NTO$1$-H; for showing
more details of the NTO$1$-E the isovalue is set to $0.03$ e/bohr$^3$. See
Table \ref{tab:Trans-Oxi}, \ref{tab:Trans-Oxi2} and the text for
more discussions.}

\label{fig:Diphenyloxirane} 
\end{figure*}

In this subsection we focus on a significantly larger system, the
trans-2,3-diphenyloxirane molecule, which contains the same CCO subsystem
as oxirane as its central motif. We show that we can use the QNTO
method to compare the excitations with those of an oxirane molecule.
As would be expected, molecules with different physical/chemical environments
can have distinct properties. For example, aryl substitution of oxirane
can lead to a different ring opening mechanism\cite{FF09}. The fact
that the trans-2,3-diphenyloxirane, referred to as the system (Sys)
molecule, is formed by two aryl substitutions on oxirane makes it
a good example for examining the influence of chemical environment
on the excitations of the core oxirane moiety using TDDFT/QNTO.

The structure of the Sys-molecule is first fully optimised without
constraints. A equivalent geometry for the oxirane molecule, referred
to as the reference (Ref) molecule, is then extracted from the optimised
Sys-molecule by replacing the two aryl substitutents with hydrogen
atoms and optimising the C-H bond lengths, at fixed bond angles (Fig. 11). The
widespread problem that the energy of charge-transfer excitations
is underestimated by local density functionals such as the LDA\cite{DH05}
means in this case that we expect to find large numbers of charge-transfer
excitations in the energy window of interest. In practice we find
we need to calculate up to $100$ excitations of the Sys-molecule
so that the highest ($100$th) excitation energy calculated, $7.74$
eV, exceeds the energy of the $8$th excitation of the Ref-molecule
($7.61$ eV), which was the highest Ref-excitation studied in the
current work.

The NTO$1$-H and NTO$1$-E orbitals for the excitations of Sys-molecule
(Sys-excitations) and Ref-molecule (Ref-excitations) are then compared
to each other to see if similar excitations can be identified between
the first $8$ Ref-excitations and the first $100$ Sys-excitations.
This is achieved by constructing, for each NTO$1$-H(E), a renormalised
orbital based on just the component of each on the local orbitals of
the $5$ atoms common to both geometries (the oxirane core).
We denote the NTO$1$-H(E) orbital of a Sys-excitation as $|\textrm{s}\rangle$,
and the corresponding core orbital cNTO$1$-H(E) in the form $|\textrm{sc}\rangle$.
Similarly, we define $|\textrm{r}\rangle$ and $|\textrm{rc}\rangle$
for the Ref-molecule. The $|\textrm{r}\rangle$ orbtitals are the NTO$1$-H(E)
orbitals of the original reference oxirane geometry, whereas $|\textrm{s}\rangle$,
$|\textrm{sc}\rangle$, and $|\textrm{rc}\rangle$ are constructed
via a similar procedure to that used in constructing the projections
of the previous section.

In particular, constructing $|\textrm{s}\rangle$ and $|\textrm{sc}\rangle$
requires a translation of the NTO$1$-H(E) from the Sys-molecule geometry
to the Ref-molecule geometry. Using NTO$1$-H as an example, the set below are constructed:
\begin{equation}
\begin{split}|\textrm{s}^{\prime}\rangle=|\phi_{\alpha;\textrm{s}\rightarrow\textrm{r}}\rangle\overline{V}_{\;1;\textrm{s}}^{\alpha}\\
|\textrm{sc}^{\prime}\rangle=|\phi_{\alpha;\textrm{s}\rightarrow\textrm{r}}\rangle\widetilde{V}_{\;1;\textrm{s}}^{\alpha}\\
|\textrm{rc}^{\prime}\rangle=|\phi_{\alpha;\textrm{r}}\rangle\widetilde{V}_{\;1;\textrm{r}}^{\alpha}
\end{split}
\label{eq:s_sc}
\end{equation}
followed by renormalisation as in Eq. \ref{eq:renormal} to obtain
$|\textrm{s}\rangle$, $|\textrm{sc}\rangle$, and $|\textrm{rc}\rangle$,
respectively.
The ``s$\rightarrow$r'' in Eq. \ref{eq:s_sc} denotes that the orbital is translated
from the Sys-molecule to the Ref-molecule.
$\widetilde{U}_{\;1}^{\alpha}$ and $\widetilde{V}_{\;1}^{\alpha}$
are defined by setting respectively the coefficients $\overline{U}_{\;1}^{\alpha}$
and $\overline{V}_{\;1}^{\alpha}$ in Eq. \ref{eq:NTOinNGWFs} to
zero for those NGWFs not centred on the core atoms; the subscript
`s' for Sys-excitations and `r' for Ref-excitations indicate that
the original NGWF coefficient matrices have been used. 

By evaluating the overlap $\langle\textrm{rc}|\textrm{sc}\rangle$ between
an NTO$n$-H or NTO$n$-E orbital of the Sys-molecule and the corresponding
NTO of the Ref-molecule, one can then tell whether there is a strong
overlap of the character in the core region.
In addition, the overlaps $\langle\textrm{sc}|\textrm{s}\rangle$ and
$\langle\textrm{rc}|\textrm{r}\rangle$ allow one to determine how
large a proportion of the total excitation is accounted for by the
component within the core region. 

In the general cases where excitations are dominated by the NTO$1$,
one can then say that the two excitations from different
molecules are similar if the following three conditions hold:
(1) a Sys-excitation must have a substantial $\langle\textrm{sc}|\textrm{s}\rangle$
for both NTO$1$-H and NTO$1$-E, (2) a Ref-excitation must have a substantial
$\langle\textrm{rc}|\textrm{r}\rangle$ for both NTO$1$-H and NTO$1$-E,
and (3) the magnitude of $\langle\textrm{rc}|\textrm{sc}\rangle$
must be large.

We find that of all the $100$ excitations of the larger system, there
are $16$ whose cNTO$1$-H and cNTO$1$-E respectively match with the cNTO$1$-H
and cNTO$1$-E of one or more Ref-excitations. We choose a threshold
of $|\langle\textrm{rc}|\textrm{sc}\rangle|\geq1/\sqrt{2}$
for both cNTO$1$-H and cNTO$1$-E. Overall there are $20$ matching
combinations, as some excitations overlap significantly with
several excitations of the reference system.
The $1$st, $2$nd, $4$th, and $8$th Ref-excitation are involved in
these matching pairs.

If the threshold is lowered to $|\langle\textrm{rc}|\textrm{sc}\rangle|\geq1/\sqrt{3}$,
the number of involved Sys-excitations becomes $25$ and the number
of matchings becomes $35$. The $3$rd and $7$th Ref-excitation also
now match with some Sys-excitations. All the $35$ excitation
matchings are listed in Table \ref{tab:Trans-Oxi}.

\begin{table}
\caption{Matchings between Sys-excitations and Ref-excitations having $|\langle\textrm{rc}|\textrm{sc}\rangle|\geq1/\sqrt{3}$
for both the NTO$1$-H and NTO$1$-E. The first and second number
in a ``(.)'' is the projection result ($\%$) for NTO$1$-H and NTO$1$-E,
respectively. $\underline{00}$ denotes $100$. Numbers shown in boldface
are the excitations plotted in Fig. \ref{fig:Diphenyloxirane}. Given
the freedom of orbital shape outside the core region, it should not be
surprising to see that the $2$nd and $4$th Ref-excitation match
with multiple Sys-excitations.}
\begin{ruledtabular}
\tiny
\begin{tabular}{c|cccccccc}

       Ref &          1 &          2 &          3 &          4 &          5 &          6 &          7 &          8 \\

$\langle \textrm{rc}|\textrm{r} \rangle$ &    (96;76) &    (96;72) &    (96;99) &    (96;78) &    (96;98) &    (96;98) &    (96;51) &   (\underline{00};75) \\
\hline
       Sys &         15 &   {\bf 34} &         26 &          6 &            &            &         15 &         16 \\

$\langle \textrm{sc}|\textrm{s} \rangle$ &    (80;38) &    (78;81) &    (79;32) &    (81;15) &            &            &    (80;38) &    (32;39) \\
$\langle \textrm{rc}|\textrm{sc} \rangle$ &  (-\underline{00};73) & (-\underline{00};-87) &  (-\underline{00};65) &   (-99;61) &            &            & (-\underline{00};-59) &   (-90;74) \\
\hline
       Sys &         24 &         43 &         39 &         14 &            &            &         24 &         21 \\

$\langle \textrm{sc}|\textrm{s} \rangle$ &    (48;39) &    (81;34) &    (51;37) &    (49;19) &            &            &    (48;39) &    (04;39) \\
$\langle \textrm{rc}|\textrm{sc} \rangle$ &  (-\underline{00};74) &   (\underline{00};84) &  (-\underline{00};68) &    (97;65) &            &            & (-\underline{00};-60) &    (80;73) \\
\hline
       Sys &   {\bf 37} &   {\bf 44} &         59 &   {\bf 34} &     {\bf } &            &   {\bf 37} &         31 \\

$\langle \textrm{sc}|\textrm{s} \rangle$ &    (81;65) &    (65;87) &    (39;43) &    (78;81) &            &            &    (81;65) &    (50;24) \\
$\langle \textrm{rc}|\textrm{sc} \rangle$ &  (\underline{00};-75) & (-\underline{00};-86) &    (79;60) &   (\underline{00};83) &            &            &   (\underline{00};63) &  (-92;-65) \\
\hline
       Sys &         52 &         58 &   {\bf (81)} &         43 &     {\bf } &            &         52 &         41 \\

$\langle \textrm{sc}|\textrm{s} \rangle$ &    (47;66) &    (13;39) &    (51;62) &    (81;34) &            &            &    (47;66) &    (32;63) \\
$\langle \textrm{rc}|\textrm{sc} \rangle$ & (-\underline{00};-75) &   (-72;81) &  (\underline{00};-58) & (-\underline{00};-76) &            &            &  (-\underline{00};63) &   (90;-78) \\
\hline
       Sys &         57 &            &     {\bf } &   {\bf 44} &            &            &         57 &         46 \\

$\langle \textrm{sc}|\textrm{s} \rangle$ &    (61;38) &            &            &    (65;87) &            &            &    (61;38) &    (04;65) \\
$\langle \textrm{rc}|\textrm{sc} \rangle$ &    (99;72) &            &            &   (\underline{00};83) &            &            &   (99;-60) &   (75;-78) \\
\hline
       Sys &         65 &            &            &         58 &            &            &         79 &         64 \\

$\langle \textrm{sc}|\textrm{s} \rangle$ &    (28;64) &            &            &    (13;39) &            &            &    (42;53) &    (29;65) \\
$\langle \textrm{rc}|\textrm{sc} \rangle$ &   (-66;68) &            &            &   (72;-72) &            &            &   (99;-61) &   (-87;62) \\
\hline
       Sys &         79 &            &            &            &            &            &            &   {\bf (71)} \\

$\langle \textrm{sc}|\textrm{s} \rangle$ &    (42;53) &            &            &            &            &            &            &    (73;37) \\
$\langle \textrm{rc}|\textrm{sc} \rangle$ &   (\underline{00};78) &            &            &            &            &            &            &    (99;74) \\
\hline
       Sys &            &            &            &            &            &            &            &         74 \\

$\langle \textrm{sc}|\textrm{s} \rangle$ &            &            &            &            &            &            &            &    (03;57) \\
$\langle \textrm{rc}|\textrm{sc} \rangle$ &            &            &            &            &            &            &            &   (-75;76) \\

\end{tabular}  
\end{ruledtabular}
~\label{tab:Trans-Oxi}
\end{table}

In Table \ref{tab:Trans-Oxi}, we see that $|\langle\textrm{rc}|\textrm{r}\rangle|\geq1/\sqrt{2}$
for both NTO$1$-H and NTO$1$-E of the $1$st-$6$th and $8$th Ref-excitation.
As for $\langle\textrm{sc}|\textrm{s}\rangle$, of all the involved
$25$ Sys-excitations discussed above only the $34$th Sys-excitation
(hereafter denoted as Sys-$34$) satisfies $|\langle\textrm{sc}|\textrm{s}\rangle|\geq1/\sqrt{2}$
for both NTO$1$-H and NTO$1$-E. In other words, for the remaining
$24$ Sys-excitations there are $\ge50\%$ density contributions of
NTO$1$-H and/or NTO$1$-E which come from outside of the core region. This
would be unsurprising as there remain $22$ atoms outside of the core
region compared to $5$ atoms in the core, and the de-localisation
of the density contribution for many Sys-excitations is anticipated.
If we lower the threshold to $|\langle\textrm{sc}|\textrm{s}\rangle|\geq1/\sqrt{3}$
for both NTO$1$-H and NTO$1$-E, the Sys-$37$ and $44$ also fall
into the set. Some excitation properties of the Sys-$34$, $37$,
and $44$ along with two more excitations, Sys-$71$ and $81$, respectively
for comparing with the Ref-$08$ and Ref-$03$, are listed in Table \ref{tab:Trans-Oxi2}
and the NTO$1$-H(E) orbitals plotted in Fig. \ref{fig:Diphenyloxirane}.
It can be seen in Table \ref{tab:Trans-Oxi2} that the NTO$1$ transition
vector component, $\sqrt{\lambda_{1}}$, has generally decreased for
the larger-sized Sys-molecule, although the NTO$1$ still accounts
for at least $60\%$ density contribution.

In Fig. \ref{fig:Diphenyloxirane} we see that the NTO$1$-H orbitals
are essentially the same for the first $7$ Ref-excitations. Moreover,
the core parts of the Sys-excitations, Sys-$34$, $37$, $44$, and
$81$, fully coincide with the core parts of this NTO$1$-H orbital
shared by the first $7$ Ref-excitations, namely, $|\langle\textrm{rc}|\textrm{sc}\rangle|=1.00$.
The differences between Sys-excitations and Ref-excitations mainly
come from the more diffuse NTO$1$-E orbitals.

Strictly speaking, only the Sys-$34$ has a significant orbital
component of NTO$1$-E coming from the core region, namely 
$|\langle\textrm{sc}|\textrm{s}\rangle|^{2}\geq0.5$.
In Table \ref{tab:Trans-Oxi2}, it can be seen that the Sys-$34$
has an excitation energy that is close to those of the Ref-$02$ and
Ref-$04$. The oscillator strength of Sys-$34$ also falls in between
those of the Ref-$02$ and Ref-$04$. As for the orbital shape, the
NTO$1$-E of the Sys-$34$ in Fig. \ref{fig:Diphenyloxirane} appears
to be a mix between those of the Ref-$02$ and Ref-$04$, also reflected
by a substantial $|\langle\textrm{rc}|\textrm{sc}\rangle|$ for both
the Ref-$02$ and Ref-$04$.

The other Sys-excitations have a high proportion of NTO$1$-H and/or
NTO$1$-E on the region outside the core, so comparing these
Sys-excitations with the Ref-excitations would be misleading.
In fact, the excitation energy can be close, as between the Ref-$04$
and Sys-$44$, or significantly different as between the Ref-$03$
and Sys-$81$. Nevertheless, the energies of similar-core excitations
between the Sys- and Ref-molecule still fall reasonably close to each
other, given the wide ranges of Sys- and Ref-excitation energies,
i.e. $4.85$-$7.74$ eV for the Sys-molecule and $6.04$-$7.61$ eV
for the Ref-molecule. Hence, the present method has been able to locate
similar excitations of the two molecules among numerous other excitations,
by appropriately tuning the thresholds of $\langle\textrm{rc}|\textrm{sc}\rangle$,
$\langle\textrm{sc}|\textrm{s}\rangle$, and $\langle\textrm{rc}|\textrm{r}\rangle$.

It would be interesting to perform a similar
study, and its ring opening reaction, using self-interaction error 
corrected or reduced functionals such as range-separated ones
\cite{CH08,LCGH11} in the near future\cite{DHS13}. 
For reducing charge-transfer contamination to local excitations, using 
some technique to confine electronic transitions to a small 
spatial range to counteract the de-localisation error from 
the LDA may also be a feasible approach.

\begin{table}
\caption{Excitation energy (E), oscillator strength (f), and NTO$1$ transition
vector component of several excitations listed in Table \ref{tab:Trans-Oxi}.
Sys-excitations and Ref-excitations that have similar core-orbitals
are put in the same row. We note that the Ref-$02$ and Ref-$04$
are substantially similar in their cNTO$1$-E orbitals to have $\langle\textrm{rc}_{\textrm{2nd}}|\textrm{rc}_{\textrm{4th}}\rangle=-0.79$.}

\begin{ruledtabular}
\begin{tabular}{c|cccc}
           &     Ref-01 &     Ref-07 &     Sys-37 &            \\

    E (eV) &       6.04 &       7.46 &       6.55 &            \\

         f &     0.0253 &     0.0114 &     0.0070 &            \\

      NTO1 &     0.9997 &     0.9998 &     0.9876 &            \\
\hline
           &     Ref-02 &     Ref-04 &     Sys-34 &     Sys-44 \\

    E (eV) &       6.52 &       6.62 &       6.44 &       6.75 \\

         f &     0.0001 &     0.0199 &     0.0062 &     0.0489 \\

      NTO1 &     0.9998 &     0.9982 &     0.9130 &     0.7740 \\
\hline
           &     Ref-03 &            &     Sys-81 &            \\

    E (eV) &       6.57 &            &       7.36 &            \\

         f &     0.0039 &            &     0.0332 &            \\

      NTO1 &     0.9998 &            &     0.8154 &            \\
\hline
           &     Ref-08 &            &     Sys-71 &            \\

    E (eV) &       7.61 &            &       7.20 &            \\

         f &     0.0028 &            &     0.0054 &            \\

      NTO1 &     0.9982 &            &     0.8851 &            \\

\end{tabular}  
\end{ruledtabular}
~\label{tab:Trans-Oxi2} 
\end{table}

\section{Conclusions and future work}

In this work we have developed QNTO analysis based on a non-orthogonal
basis, as implemented in the linear-scaling DFT/TDDFT code ONETEP.
We use the QNTO method to analyse the transition vectors of electronic
excitations of various molecular systems offered by TDDFT calculations.

We first studied the photo-induced ring-opening process of
the oxirane molecule. Through a ground state scan along the reactive CCO angle of
an oxirane molecule followed by TDDFT calculations (vertical excitations), 
a profile of several
low-lying adiabatic excited state curves on which the excited state
driven reaction may be carried can be constructed. However, these
adiabatic excited state curves ranked by energies do not give us very
much knowledge on how the reaction will proceed. We stress that the
reaction process can be made much clearer if the excited state potential
energy curves are labelled according to their excitation character
instead of their energies. We demonstrated that the transition
vectors of these low-lying excitations are dominated by the first
NTO electron-hole pair (NTO$1$), and therefore the transition origins
of NTO$1$ can be used to represent the character of these excitations.
Based on this approach, we have successfully located several special
geometrical points along the scanned CCO coordinate where the transition
origins of two adiabatic excited state curves switch. By reconnecting
the adiabatic excited state energy points (calculated by TDDFT) based
on transition origins instead of energetic ordering, we are then
able to depict how the excited state-driven ring-opening process
can proceed. This picture is consistent with the previous excited
state molecular dynamics study that found that the population of the
$^{1}$(n,$3p_{z}$) state facilitates the ring-opening process, while
the population of $^{1}$(n,$3s$) state does not. We have demonstrated
that the $^{1}$(n,$3p_{z}$) state exhibits a monotonically-decreasing
energy curve along the widening CCO angle, leading to a ring-opening
reaction. For more complicated reactions with several coordinates 
involved and geometry strongly distorted, more geometries
may be used, e.g. those from an excited state scan and molecular 
dynamics simulation, to help identify the photo-driven 
reaction pathways of interest. A promising future direction
could be to apply this methodology to larger molecules with much more
complex reaction pathways.

QNTO analysis has also been employed to study an aryl-substituted
oxirane, the trans-2,3-diphenyloxirane molecule. The results demonstrate
that the approach can be used to identify specific excitations among a
large number of other excitations of a large/complex molecular system.
By comparing the excitations of this larger molecule to those of oxirane,
we have shown that a small number of mutually similar excitations
can be identified and compared between the two molecules. The
matching conditions that define similar excitations can be tuned by
the magnitudes of: (1) the projection result of the core orbitals,
cNTO$1$-H(E), between Sys-excitations and Ref-excitations; (2) the
proportion of the cNTO$1$-H(E) accounting for the whole NTO$1$-H(E)
in the Sys-excitation; (3) the proportion of the cNTO$1$-H(E)
accounting for the whole NTO$1$-H(E) in the Ref-excitation. For example,
a Sys-excitation can have a similar cNTO$1$-E to that of a Ref-excitation,
but the cNTO$1$-E of the Sys-excitation only accounts for a tiny
proportion to the whole NTO$1$-H(E) of Sys-excitation. In such a
case, the orbital-section of the NTO$1$-E of Sys-excitation coming from outside of
the core region cannot be ignored and the two excitations can not
be assigned as similar excitations. The same approach for finding
similar excitations through the transition origins would be particularly
powerful when there are numerous excitations in a large molecular
system. For example, those excitations that are located in a specific
geometric region can be readily identified, helping to analyse the
chemical properties of that region in a specific energy range of light
absorption.

\section{Acknowledgement}

The authors would like to thank Louis Lee for very helpful discussions.
All calculations in this work were carried out using the Cambridge
HPC Service under EPSRC Grant EP/J017639/1. JHL acknowledges the support of Taiwan Cambridge Scholarship. TJZ acknowledges
the support of EPSRC Grant EP/J017639/1 and the ARCHER eCSE programme.
NDMH acknowledges the support of the Winton Programme for the Physics
of Sustainability. 

\newpage{}

\bibliographystyle{apsrev4-1}

\end{document}